\documentclass[traditabstract]{aa}

\usepackage{graphicx}
\usepackage{subfigure}
\usepackage{multirow}
\usepackage{color}
\usepackage[usenames,dvipsnames]{xcolor}
\usepackage{natbib}
\usepackage{amsmath}
\usepackage{bbold}
\usepackage{scalefnt}
\usepackage[T1]{fontenc}
\usepackage{mathtools}
\bibpunct{(}{)}{;}{a}{}{,}
\usepackage{caption}
\usepackage{hyperref}
\usepackage{threeparttable}

\newbox\grsign \setbox\grsign=\hbox{$>$} \newdimen\grdimen \grdimen=\ht\grsign
\newbox\simlessbox \newbox\simgreatbox \newbox\simpropbox \newbox\wtildebox 
\setbox\simgreatbox=\hbox{\raise.5ex\hbox{$>$}\llap
     {\lower.5ex\hbox{$\sim$}}}\ht1=\grdimen\dp1=0pt
\setbox\simlessbox=\hbox{\raise.5ex\hbox{$<$}\llap
     {\lower.5ex\hbox{$\sim$}}}\ht2=\grdimen\dp2=0pt

\newcommand{\Msun}{\mbox{$M_{\odot}$}}

\newcommand{\be}{\mbox{\begin{equation}}}
\newcommand{\ee}{\mbox{\end{equation}}}

\newcommand{\Cref}{\mbox{$m_{\rm ref}$}}

\begin{document}

\title{Imaging diagnostics for transitional discs}

\author{M.\,de\,Juan\,Ovelar\inst{1}, M.\,Min\inst{2}, C.\,Dominik\inst{2,3}, C.\,Thalmann\inst{2}, P.\,Pinilla\inst{4}, M.\, Benisty\inst{5} and T. Birnstiel\inst{6}}

\institute { 
                 {\inst{1}Leiden Observatory, Leiden University,
                 P.O. Box 9513, 2300RA Leiden, The Netherlands}\\
                 {\inst{2}Astronomical Institute Anton Pannekoek, 
                 University of Amsterdam, 1090 GE Amsterdam, The Netherlands}\\
                 {\inst{3}Department of Astrophysics/IMAPP, Radboud 
                 University Nijmegen, 6500 GL Nijmegen, The Netherlands}\\
                 {\inst{4}Institut f\"{u}r Theoretische Astrophysik, 
                 Universit\"{a}t Heidelberg, Albert-Ueberle-Stra\ss e 2, 69120 Heidelberg, Germany}\\
                 {\inst{5}Institut de Plan\`{e}tologie et Astrophysique Grenoble, 
                 414 rue de la Piscine, 38400 st. Martin d'Heres, France}\\
                 {\inst{6}Harvard-Smithsonian Center for Astrophysics, 
                 60 Garden Street, Cambridge, MA 02138, USA}
                 }
\date{Received date; accepted date}

\abstract{{Transitional discs are a special type of protoplanetary discs where planet formation {is thought to} be taking place}. {These objects feature characteristic inner cavities and/or gaps of a few tens of AUs in the sub-millimetre images of the disc. This signature suggests} a localised depletion of matter in the disc that could be caused by {planet formation processes}. {However, recent observations have revealed {differences in} the structures imaged at different wavelengths in some of these discs}. In this paper, we aim to explain these {observational differences {using} self-consistent physical 2-D hydrodynamical and dust evolution models of such {objects}, {assuming their morphology is indeed generated by the presence of a planet}. We use these models to derive the {distribution of gas and dust} {in a {theoretical} planet-hosting disc, for various planet masses and orbital separations}. We then simulate observations of the emitted and scattered light from these models with VLT/{SPHERE ZIMPOL}, Subaru/HiCIAO, VLT/VISIR and ALMA. We do this by first computing the full resolution images of the models at different wavelengths, and then simulating the observations accounting for the characteristics of each particular instrument}. The presence of the planet generates pressure bumps in the gas distribution of the disc {whose characteristics strongly depend on the planet mass and position}. These bumps cause large grains to accumulate while small grains are allowed into inner regions. This spatial differentiation of the grain sizes explains the {differences} in the observations since different wavelengths and observing techniques trace different parts of the dust size distribution. Based on this effect, we conclude that the combination of visible/near-infrared polarimetric and sub-mm images is the best strategy to constrain the properties {of the unseen planet responsible for the disc structure}.}

\keywords{
Transition Discs -- Protoplanetary discs -- (Stars:) circumstellar matter --
}

\authorrunning{M.\,de\,Juan\,Ovelar et al.}
\titlerunning{Imaging diagnostics for transitional discs}

\maketitle

\section{Introduction} \label{sec:intro}

Transitional discs are {generally} believed to be the result of a planet forming stage in {a circumstellar disc.} {As such, their study provides with important information that can help us to constrain the physics of the planet formation process}. Therefore, a considerable amount of effort is currently being invested {to understand} these objects {better}, both theoretically and observationally. {In general, the spectral energy distribution (SED) of these sources as well as interferometric measurements at sub-millimitre wavelengths show evidence for inner cavities and/or gaps \citep[e.g.][]{strom89,espaillat10,andrews11}. }
{One of the interpretations} of these observations is that a fraction of the material in the disc is {{depleted} by} a forming planet \citep[see the review of][for an extensive view]{williams11}.  

{Recently} observational work has {called this conclusion into question} based on e.g.\,measurements of the accretion rate \citep[e.g.][]{calvet05,espaillat07} and polarimetric observations of the inner regions of the disc \citep{dong12}. The latter one, of particular interest since it involves direct imaging of the disc structure, was carried out as a part of the near infrared (NIR) Strategic Explorations of Exoplanets and Disks with Subaru survey (SEEDS, \citealt{tamura09}). The results of the survey were surprising regarding a number of transition discs classified as such {based on} sub-mm emission observations. The polarimetric images {of these discs} did not show the expected gap \citep{dong12}. A parametric model of the disc with a continuous radial distribution of small grains $\sim1\,\rm{\mu m}$ and a significant depletion of big grains $\sim1\,\rm{mm}$ in the inner regions of the disc was found to reproduce such observational discrepancies. These regions would appear empty in the sub-mm images whereas the presence of small grains, scattering light very efficiently in the NIR wavelengths, would explain the polarimetric measurements. 
However, due to the parametric nature of the model, the {\citet{dong12} study} could not provide an explanation of the physical mechanisms causing this differentiation of big and small dust grain distributions across the radial extent of the disc, but merely suggested that some filtering mechanism must be {active}.

Theoretical studies such as \citet{rice06,zhu11, zhu12,pinilla12a} provide a potential physical explanation. The main drive for these studies is to explain how particles of dust in a disc can grow up to large sizes ($\sim1\,\rm{m}$) without being dragged towards the star by the radial drift mechanism. This constitutes one of the long standing problems in planets formation, the \emph{``$1$-m barrier''} problem \citep{weidenschilling77,brauer08a}. In particular, one of the latest studies \citep{pinilla12a}, explored the influence the presence of a planet has on the distribution of gas and dust in the disc. In this study, the authors combine 2-D hydrodynamical simulations of the evolution of the gas distribution in a disc that hosts a planet with {state of the art} dust evolution models. {These models include, for the first time,} self-consistent calculations of the radial drift, coagulation and fragmentation mechanisms undergone by the dust \citep{birnstiel10a}. They showed that the presence of a planet generates pressure gradients in the gas distribution that cause the velocity of the gas to increase to near-Keplerian values at specific radial positions. This reduces the differential velocity between dust and gas particles and, therefore, the drag force exerted on the dust. As a consequence, large particles ($\sim 1\,\rm{mm}$), {which are less affected by the gas drag, accumulate} in those radial locations where the pressure reaches a maximum, which allows them to grow ``protected'' from the radial drift. Smaller grains, still coupled to the gas, {follow the accretion flow into} the inner regions of the disc. The trapping and filtering characteristics and the radial position of these bumps are highly dependent on the mass and position of the planet.

If these models indeed reproduce the physical processes taking place in a transitional disc, measuring the characteristics of this size-differentiated dust distribution {should allow} to constrain characteristics of the planet that causes it. The observational {differences} found between sub-mm emission and NIR scattered images become then a powerful diagnostic tool for transitional discs and the planets they host. In this work, we {perform a theoretical study simulating} observations of {hydrodynamical and dust evolution models, similar to those presented in \citet{pinilla12a},} with {SPHERE ZIMPOL}, HiCIAO, VISIR and ALMA. {Our aim is} to 1) test whether {these models can reproduce the general characteristic features} found in observations, 2) analyse what different imaging techniques can tell us about the dust distribution and 3) detect the best imaging strategy to constrain characteristics of the planet such as mass and position from dust measurements. {We would like to clarify that we do not aim to explain detailed characteristics revealed by particular observations of transitional discs.}

The study is organised as follows: In Section\,\ref{sec:method}, we describe the methodology used to obtain the disc models and the simulated observations. In Section\,\ref{sec:results} we present images and radial profiles obtained for the different cases and instruments considered and in Section\,\ref{sec:discussion} the discussion of those results. Finally, in Section\,\ref{sec:conclusions} we provide a short summary of the study and list our conclusions.

\section{Method} \label{sec:method}

In order to generate images of a transitional disc, we base our study in the disc-planet interaction models presented in \citet{pinilla12a}. These models combine two dimensional hydrodynamical and dust evolution simulations including radial drift, coagulation and fragmentation, to self-consistently reproduce the evolution of the gas and dust in the disc. We consider the cases of a disc with a planet of masses $M_{\rm{p}}=[1,9,15]\,{M_\mathrm{Jup}}$ {at radial positions $R_{\rm{p}}=[20,40,60]\,\rm{AU}$}. We take the resulting distribution of dust and gas for these three cases after $3\,\rm{Myr}$ of evolution and input them in the Monte-Carlo radiative transfer code MCMax, to produce {full resolution} intensity and polarised intensity images of the emitted and scattered flux. Finally, we simulate realistic observations with VLT/{SPHERE ZIMPOL}, Subaru/HiCIAO, VLT/VISIR and ALMA either using specific instrument simulators or convolving the {full resolution} images with the characteristic {point spread function} (PSF) and adding realistic effects such as noise, and decrease in resolution due to seeing. In the following subsections we present the details of the steps followed.

\subsection{Disc-planet interaction models}\label{subsec:models}
\begin{figure*}[!htb]
   \centering
   \begin{tabular}{c} 
   \includegraphics[width=5.5cm]{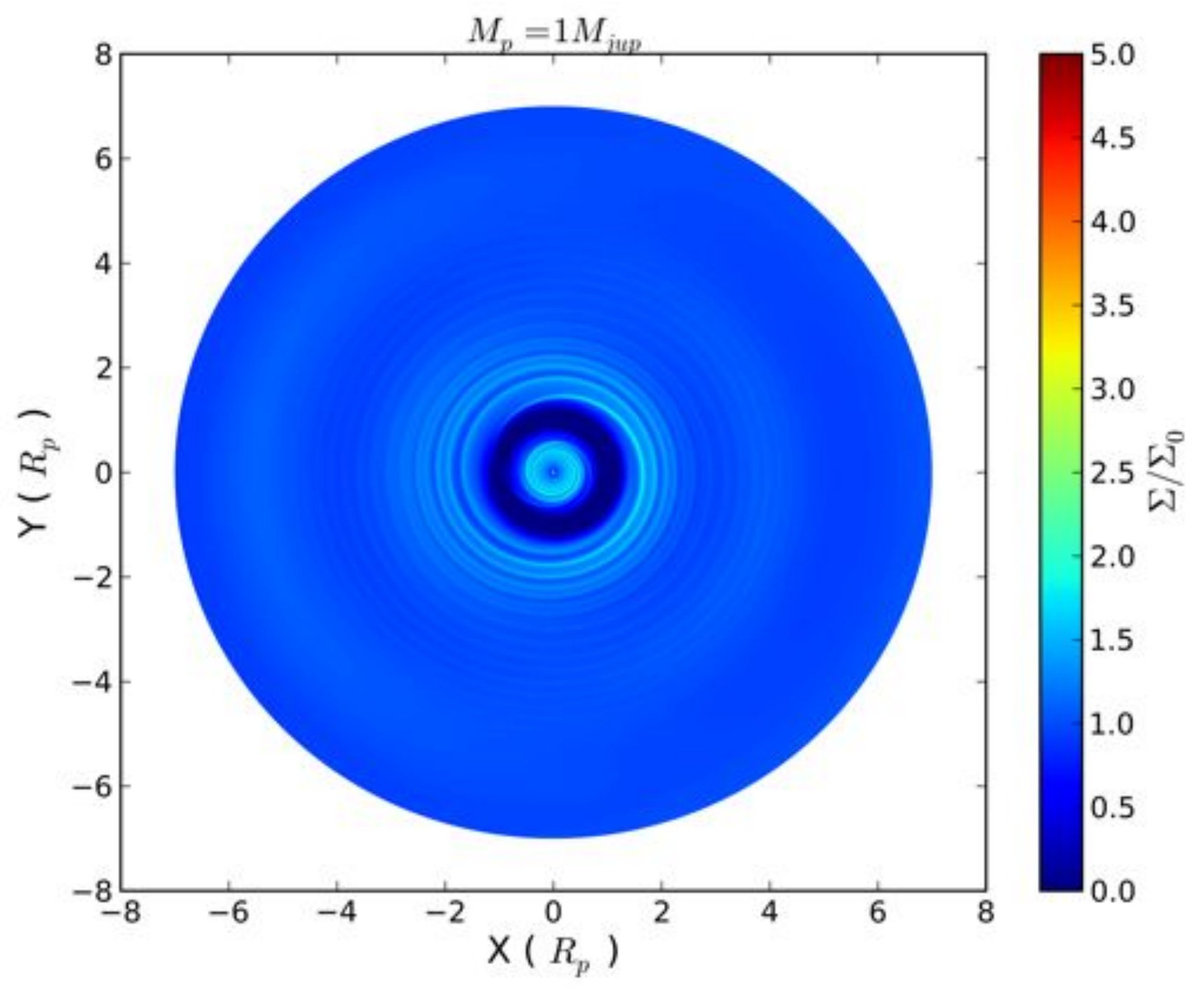}\hspace{0.8mm} 
   \includegraphics[width=5.5cm]{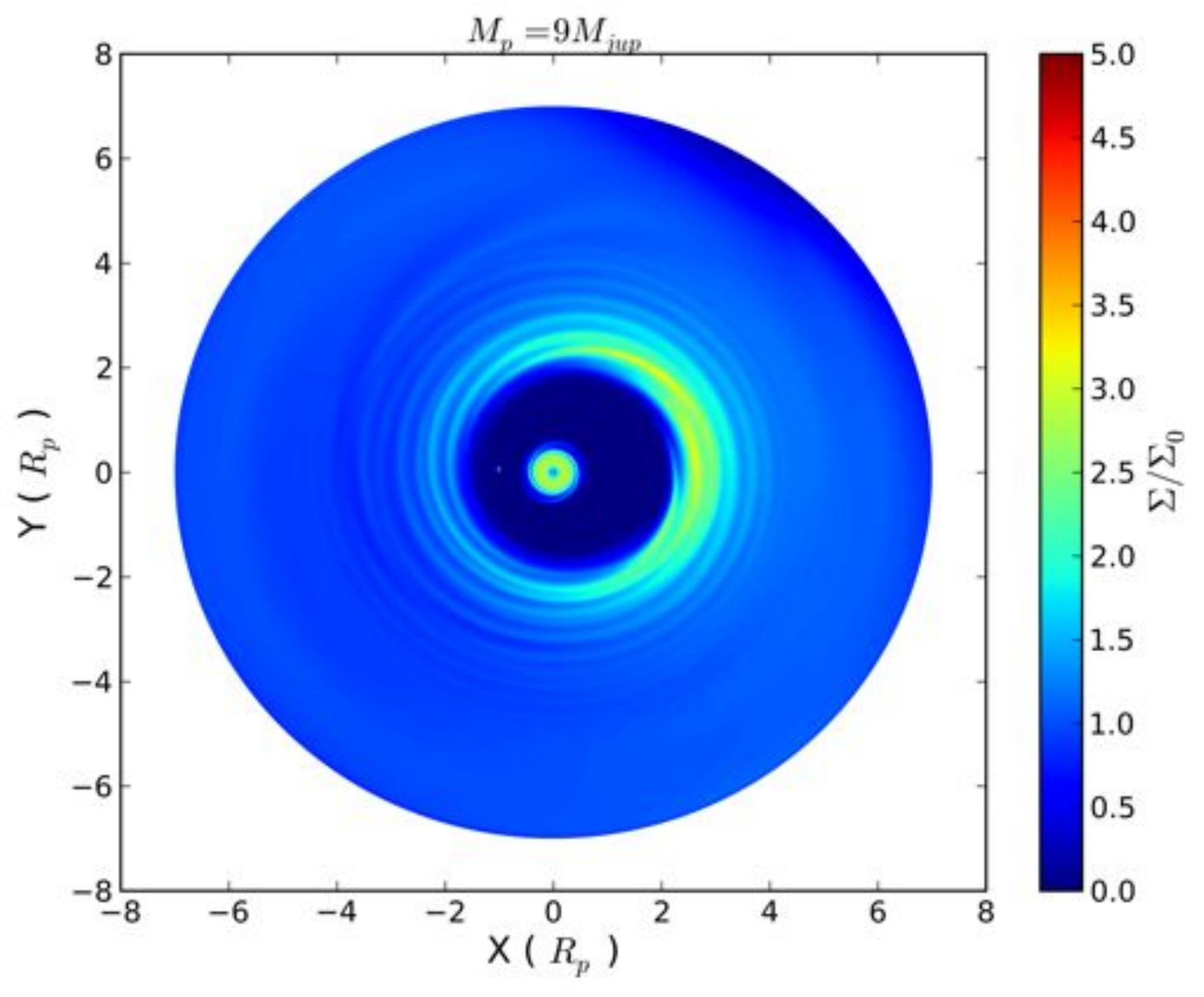}\hspace{0.8mm} 
   \includegraphics[width=5.5cm]{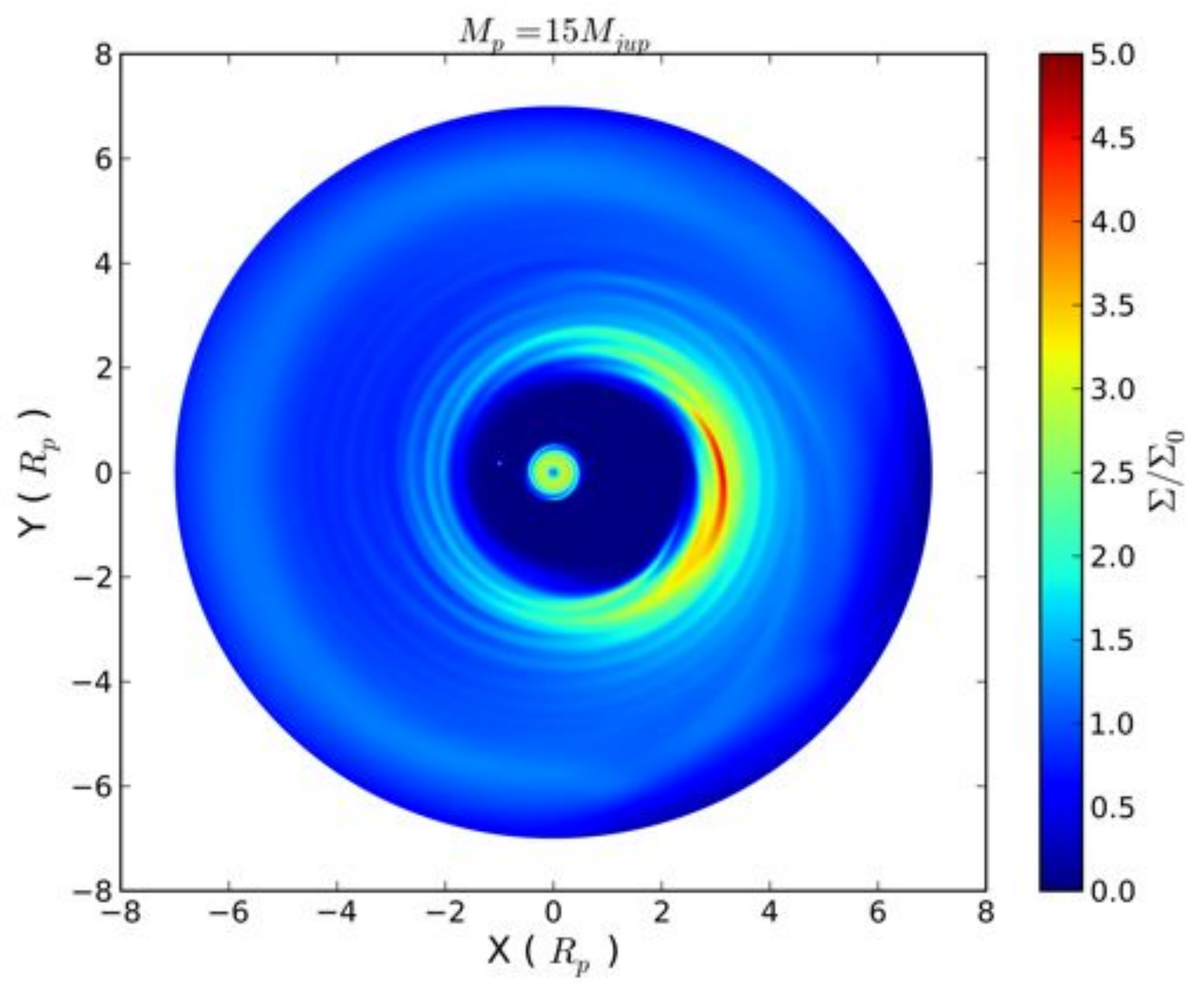}\\
   \includegraphics[width=5.5cm]{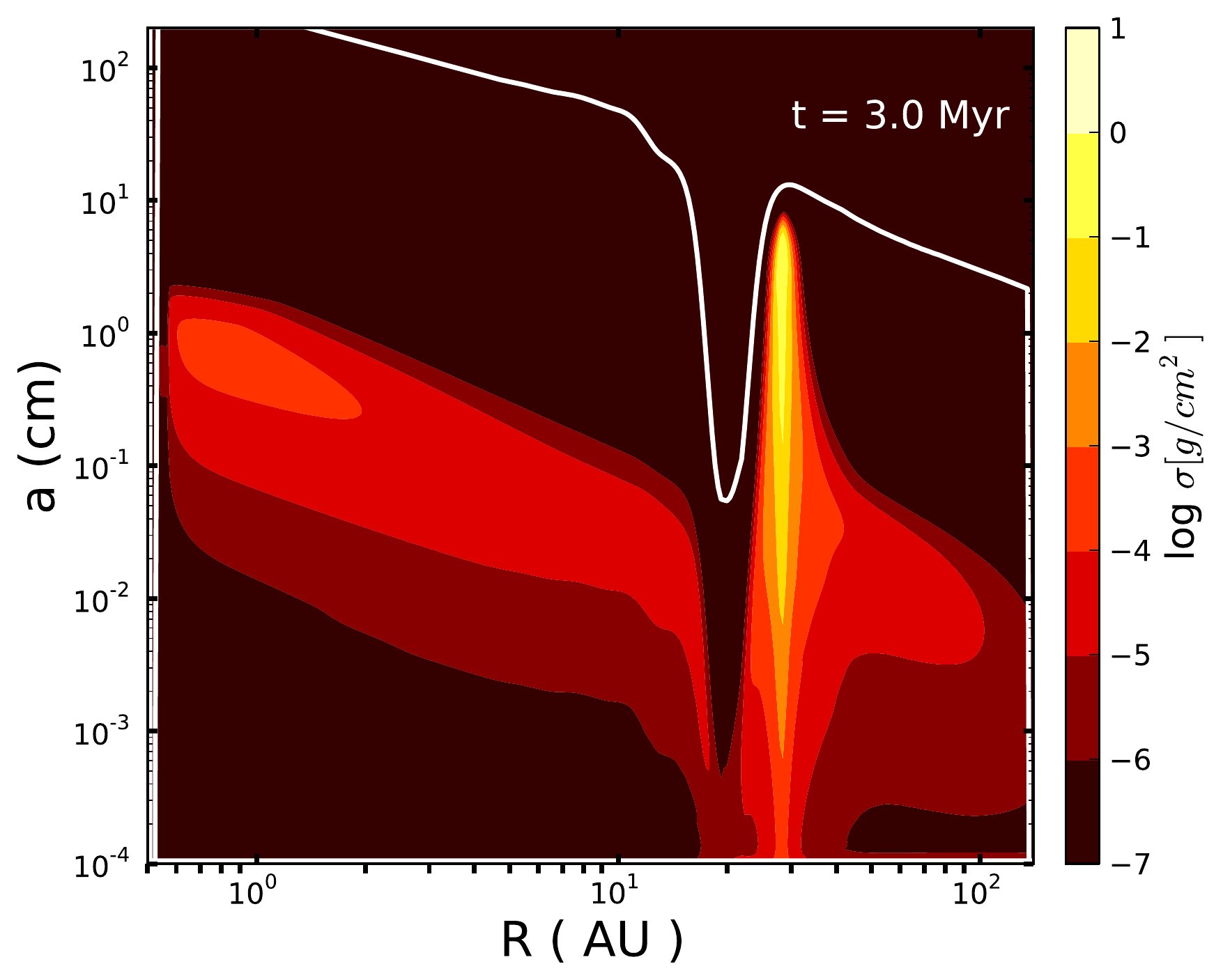} \hspace{0.3mm} 
   \includegraphics[width=5.5cm]{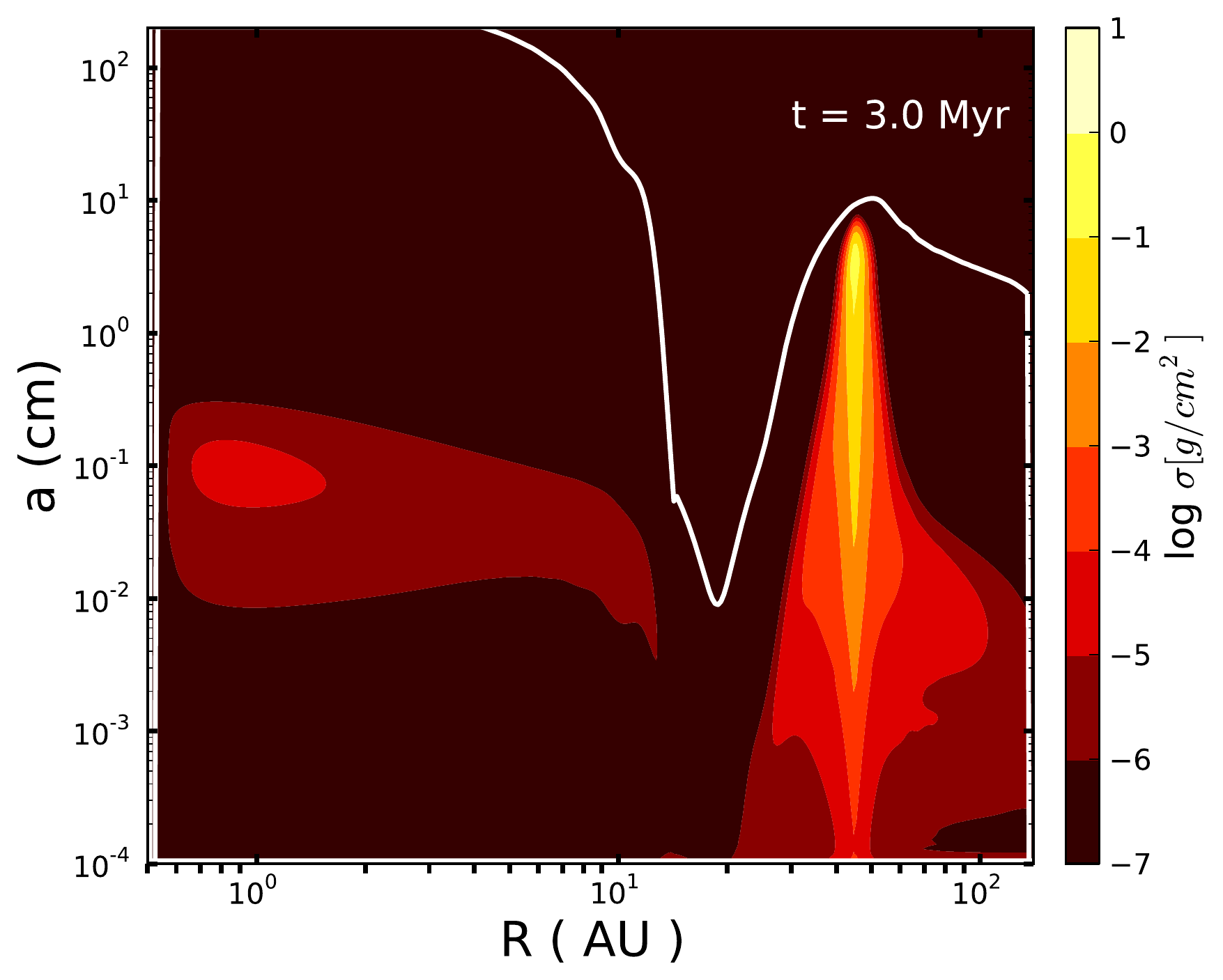}\hspace{0.3mm} 
   \includegraphics[width=5.5cm]{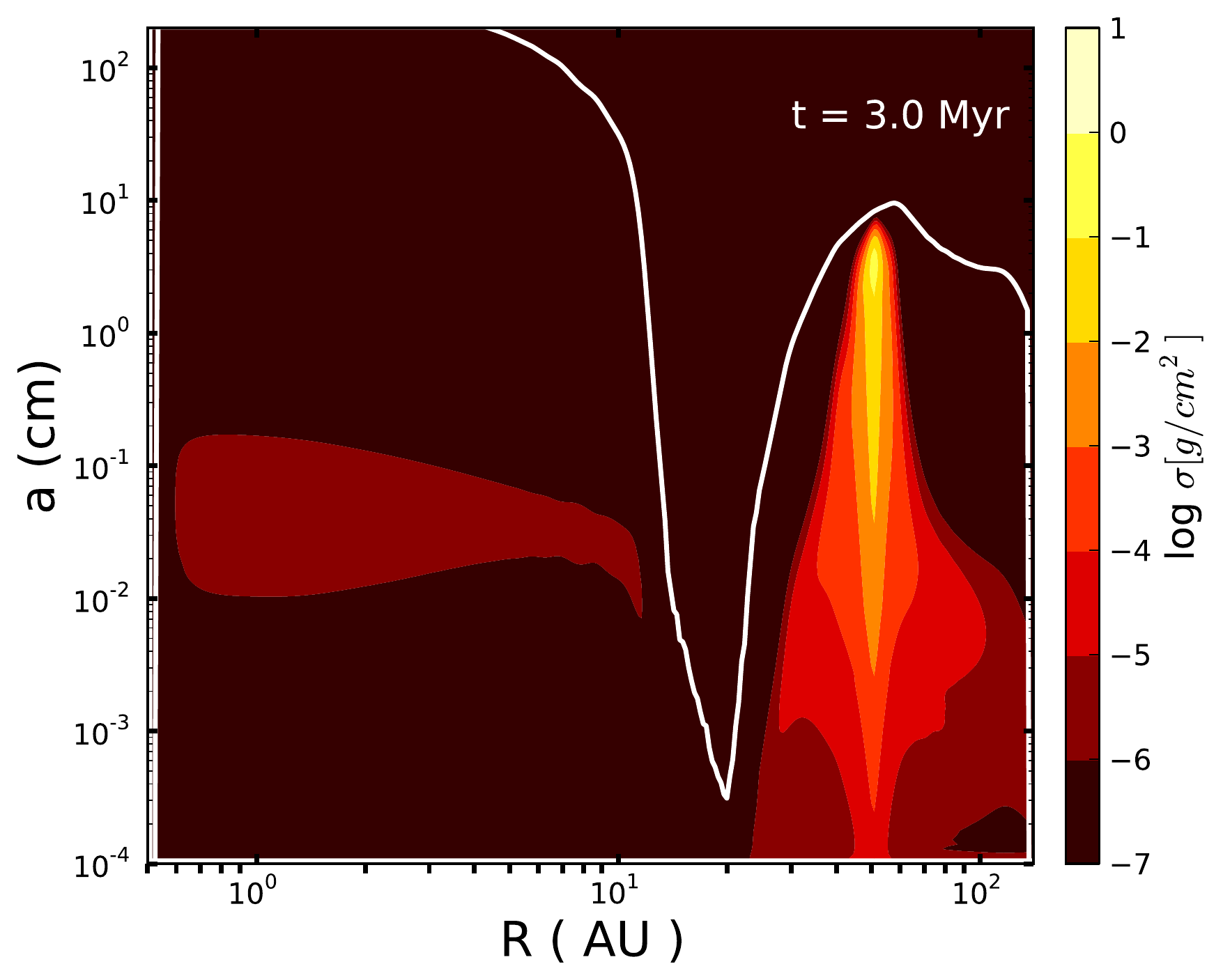}
    \end{tabular}	
   \caption{\emph{Top panels:}  2D gas surface density for $[1,9,15]\,\rm{M_{jup}}$ after 1000 orbits of evolution.  \emph{Bottom panels:} Vertically integrated  dust  density distribution for each planet mass and for the case of $R_{\rm{p}}$=20\,AU. The white line corresponds to the size of particles that feel the highest radial drift and it is proportional to the gas surface density.}
   \label{2D_1D_Hydro_dust}
\end{figure*}

\begin{table}[t]
	\renewcommand{\arraystretch}{1.3}
	\renewcommand{\captionfont}{\scshape}
	\small	
	\begin{center}
		\caption{disc-planet simulation \\ input parameters from Pinilla et al. 2012 }\label{tab:discmodel}
		\begin{tabular}{ll}
			\hline
			\hline
			Temperature of the star ($T_{\rm{star}}$) 			& $4730\,\rm{K}$\\
			Radius of the star ($R_{\rm{star}}$)				& $1.7\,\rm{\Msun}$\\
			Mass of the star ($M_{\rm{star}}$)					& $1\,{\Msun}$\\
			Mass of the disc ($M_{\rm{disc}}$)					& $0.0525\,\Msun$\\
			Position of the planet ($R_{\rm{p}}$)				& $[20,40,60]\,\rm{AU}$\\
			Fragmentation velocity ($v_{\rm{f}}$)				& $10\,\rm{m/s}$\\
			Inner disc radius ($R_{\rm{disc,inn}}$)				& $0.025\,R_{\rm{p}}$\\
			Outer disc radius ($R_{\rm{disc,out}}$)				& $7.0\,R_{\rm{p}}$\\
			Solid density of dust particles ($\rho_{\rm{dust}}$)	& $1.2\,\rm{g/cm^3}$\\
			Alpha viscosity ($\alpha$)							& $10^{-3}$\\
			\hline
		\end{tabular}
	\end{center}
\end{table}%

{To obtain the gas and dust distribution of our transitional discs we follow the same simulation procedure used in \citet{pinilla12a}. The results} are obtained {by} first solving the equations for the hydrodynamical interactions between the gas in the disc and the planet, and then computing the evolution of the dust in a disc with the obtained gas surface distribution. The first computation is done by means of the two-dimensional hydrodynamical code FARGO \citep{masset00}, which uses finite differences to solve the Navier-Stokes and continuity equations in a grid of annular cells that define the disc. The simulations were computed selecting open boundary conditions to allow the material to leave the grid. The logarithmically extended radial grid is taken for each case from $R_{\rm{disc,inn}}=0.025$ to $R_{\rm{disc,out}}=7.0$,  using normalized units, such that the planet is located at $R_{\rm{p}}=1.0$. Table\,\ref{tab:discmodel} lists the input parameters used in these simulations. {They} are the same as in \citet{pinilla12a} but considering only $\Sigma\propto r^{-1}$, kinematic viscosity  $\nu = \alpha c_s h$, with $\alpha=10^{-3}$, and normalizing the mass of the disc to $M_{\rm{disc}}=0.55\,\rm{\Msun}$. {Stellar parameters are those typical of T-Tauri stars. Note that parameters defining the geometry of the disc, i.e.\,$R_{\rm{disc,inn}}$ and $R_{\rm{disc,out}}$ are defined with respect to the position of the {planet such that they are always far enough from it allowing for a appropriate study of the effects it has in the inner and outer discs. Although this link may not be necessarily physical, it allows for comparison of the different planet separation cases without increasing the radial resolution of the simulations i.e. the required computational time.}

The hydrodynamical simulations are done until the disc reaches a quasi-stable state after {$\sim1000$} planet orbits. The 2D gas surface density is then azimuthally averaged and used as the initial condition for the dust evolution modelling. Note that during the dust evolution modelling the gas surface density remains constant {on timescales of millions of years}, hence any mechanism that may disturb the gas density as, for instance, photoevaporation, is omitted. For the dust evolution simulations, we use the {1-D} code described in \citet{birnstiel10a}, {which} computes the growth and fragmentation happening in the {radial dust distribution} due to radial drift, turbulent mixing and gas drag forces. The dust is initially distributed such that the dust-to-gas ratio is $1\%$ and with {an} initial size of $1\,\mu$m. {The dust distribution is evolved for} $3\,\rm{Myr}$. We follow this process for all planet mass and radii cases, {i.e.,}\ $M_{\rm{p}}=[1,9,15]\,{M_\mathrm{Jup}}$ and $R_\mathrm{p}=[20,40,60]\,\rm{AU}$. 

Fig.\,\ref{2D_1D_Hydro_dust} shows the results of these simulations for each planet mass studied and for the specific case of $R_{\rm{p}}=20\,\rm{AU}$. Top panels show the 2D gas surface density after 1000 orbits of evolution. For each case, it is clear how the shape of the gap varies with the mass of the planet. For the case of $15\,\rm{M_{jup}}$, even a vortex appears at the outer edge of the gap due to the high mass of the planet \citep{ataiee13}. 

Bottom panels of Fig.\,\ref{2D_1D_Hydro_dust} show the vertically integrated dust density distribution \citep[see][Eq. 5 and 6]{pinilla12a}. The solid white line indicates the size of particles that feel the highest radial drift, which is directly proportional to the azimuthally averaged gas surface density \citep[see][for details]{pinilla12a}. The presence of the planet clearly perturbs the gas and dust density distributions, dividing the disc into inner and outer regions, where the distribution of dust particle sizes is clearly different. {Note that, due to the 1-D nature of the dust evolution code, the dust distribution cannot reproduce the asymmetries caused by the presence of e.g.\,vortices in the gas distribution. For the purposes of our study, where the aim is to analyse the radial morphology of the dust distribution, this approximation is sufficient.}

\subsection{Radiative Transfer simulations}

To compute the emitted and scattered flux of the disc we use the Monte-Carlo radiative transfer code MCMax \citep{min09}. MCMax self-consistently solves the temperature and vertical structure of the distribution of gas and dust in the disc given the size and composition of the particles and the characteristics of the central star, {including the effect of dust settling}. It produces common observables for the study of the disc such as SED or emission and scattering images, at the desired wavelengths and inclination angles. MCMax is widely used to compare theoretical models of discs with observations \citep[see][for some examples]{mulders10,mulders11,mulders12,devries12,canovas12, jeffers12,lombaert12,honda12,min12,mulders13}.

{MCMax reads in the distribution of gas and dust obtained from the disc-planet models for all planet mass and {separation} cases as well as the central star and general disc parameters. It also requires a composition of the dust to compute the opacities and temperatures. We model this composition} as a mixture of silicates ($\sim$$58\%$), iron sulphide ($\sim$$0.18\%$) and carbonaceous dust grains ($\sim$$0.24\%$) with an average density of $\rho=3.2\,\rm{g/cm^{3}}$ \citep{min11}. {We set a porosity for the dust grains of $p=0.625$ which corresponds to the $\rho=1.2\,\rm{g/cm^{3}}$ used in the dust evolution simulations}. The indexes of refraction needed to compute the opacities were obtained from \citet{dorschner95,henning96} for the silicates, from \citet{begemann94} for the iron sulphide and from \citet{preibisch93} for the carbonaceous dust grains.
{MCMax self-consistently simulates the settling effect provided the viscous turbulence ($\alpha$). We set this value to that of the disc-planet simulations (i.e.\,$\alpha=10^{-3}$).} {The vertical structure of the gas in the disk is solved iteratively under the assumption of vertical hydrostatic equilibrium. The vertical structure of the dust is then computed using settling and vertical turbulent mixing.}

With the resultant temperature and vertical structure, MCMax produces intensity and polarised intensity theoretical images of the disc at the desired wavelength. In this paper we will discuss the results obtained for wavelengths of $\lambda=[0.65,1.6,20,850]\,\rm{\mu m}$ which are the most commonly used for imaging diagnostics of circumstellar matter.

\subsection{Simulated observations}

To simulate realistic observations of the modeled discs, we select a set of currently available {(Subaru/HiCIAO and VLT/VISIR)} and near future instruments {(VLT/SPHERE-ZIMPOL and ALMA Complete Array)} that are (or are likely to be) leading the field of imaging circumstellar environments. These instruments work in different regions of the electromagnetic spectrum, thus probing different features in the disc. We select a filter from the available ones in each instrument {($RI$, $H$, $Q$ and 850\,$\mu$m)}, and produce an MCMax {full-resolution} image for a particular wavelength in the filter range ($\lambda=[0.65,1.6,20,850]\,\rm{\mu m}$ respectively). Then we either process these images using a specific instrument simulator, convolving it with {a measured or simulated PSF of the instrument (depending on the case), and} adding photon noise and loss of resolution accounting for realistic observational effects. We assume an exposure time of $t_{\rm{obs}}=1200\,\rm{s}$ with all instruments. Table\,\ref{tab:resolution} shows the the theoretical and final resolution obtained for the simulated observation with each instrument. In the following paragraphs we explain the details of the image processing followed in each case.  

\begin{table}[t]
	\renewcommand{\arraystretch}{1.3}
	\renewcommand{\captionfont}{\scshape}
	\small	
	\begin{center}
		\caption{Wavelength and resolution of the\\ simulated images}\label{tab:resolution}
		\begin{tabular}{llll}
			\hline
			\hline
			Instrument			& $1.2\lambda/D$	& final		& $d=140\,\rm{pc}$\\
			\hline
			{SPHERE ZIMPOL}	& $0.02\arcsec$			&$0.03\arcsec$	&$\sim\phantom{1}4\,\rm{AU}$\\
			{Subaru} HiCIAO	& $0.04\arcsec$			&$0.09\arcsec$	&$\sim13\,\rm{AU}$\\
			{VLT} VISIR		& $0.6\arcsec$				&$0.62\arcsec$	&$\sim87\,\rm{AU}$\\
			ALMA complete		& $0.013\arcsec$			&$0.015\arcsec$	&$\sim\phantom{1}2\,\rm{AU}$\\
			\hline
		\end{tabular}
	\end{center}
\end{table}

\begin{itemize}
	\item{\emph{{$R$-band} intensity and polarised intensity images with {SPHERE ZIMPOL}}:\\
\\
Intensity and polarised intensity observations with SPHERE in {$R$}-band are simulated with the {SPHERE ZIMPOL} simulator that comes as part of the SPHERE software package \citep{thalmann08}. {SPHERE} \citep{beuzit06} is the planet finder designed for the Very Large {Telescope, {which is planned to see first light by the end of 2013}. {By means of its polarimeter ZIMPOL it provides linear polarimetric imaging} capabilities for the characterisation of circumstellar environments and exoplanets  \citep{gisler04,stuik05,thalmann08,roelfsema10,schmid10}.} The simulator takes full resolution intensity and polarimetric images as the input, and generates the observed images simulating the artifacts and aberrations caused by the optical system. 
{We} choose the RI filter in the simulator and process the MCMax intensity, Stokes $Q$ and $U$ images. The final polarised intensity image was obtained as $PI=\sqrt{Q^2+U^2}$.}\\
		
	\item{\emph{{$H$-band} intensity and polarised intensity images with HiCIAO}:\\
\\
In the case of {$H$-band} images, we convolve the full resolution intensity and Stokes $Q$ and $U$ images with a measured HiCIAO {$H$-band} PSF. This PSF was obtained from the publicly available ACORNS-ADI data reduction pipeline \citep{brandt13}. We add photon noise to the convolved images as follows. First we add photon noise to the convolved intensity image. Then the previous noise-free intensity image is subtracted from this one to generate a photon noise map. We do this two consecutive times to generate different maps for the $Q$ and $U$ images. Each map is divided by two, assuming that half of the observing time goes to each linear polarisation measurement (i.e.\,$Q$ and $U$), and the result is added to the convolved $Q$ and $U$ images. The final polarised intensity image is then computed again as $PI=\sqrt{Q^2+U^2}$. The final resolution obtained in this case is determined by the full width half maximum (FWHM) of the measured PSF.}\\
	
	\item{\emph{Q-band intensity images with VISIR}:\\
\\
To simulate {$Q$}-band observations we use MCMax. Provided the dimensions of the primary and secondary mirrors and the exposure time, the code can generate a theoretical PSF to be convolved with the images and compute the corresponding photon noise. There is also the option to specify an angular width to be added to the simulated resolution determined by the PSF. This accounts for the effect of the observing conditions ({i.e.,}\ seeing). We specify values for the primary and secondary mirrors of $D_1=8.4\,\rm{m}$ and $D_2=1\,\rm{m}$, an exposure time of $t_{\rm{obs}}=12000\,\rm{s}$ and a seeing width of $w=0.05\arcsec$.}\\
	
	\item{\emph{$850\,\rm{\mu m}$ intensity images with ALMA}:\\
\\
{The ALMA observations are simulated in a simplified way. The spatial resolution of ALMA can be estimated by calculating the resolution of a telescope with a primary mirror diameter as big as the baseline of the {antenna} array. For our study, we assume the maximum baseline provided by ALMA complete, which corresponds to $16\,\rm{km}$, and the diffraction-limited resolution at $850\,\rm{\mu m}$ is $13\,\rm{mas}$. {The ALMA complete array was assembled and inaugurated in March 2013, although some antennas are still being tested. Observations for the next cycle (Early Science-Cycle 2), with still reduced capabilities, are expected to be performed in mid-2014. Full operations are expected to follow shortly after the end of Cycle 2 operations}. To allow for slightly reduced performance, we use $15\,\rm{mas}$ as the FWHM of the PSF. We adjusted the exposure time to account for the fact that the actual collecting area of ALMA is less than that of a telescope with a mirror as big as the baseline. In the final configuration of ALMA complete (using only the $12\,\rm{m}$ {antennas}) the collecting area will be provided by $50$ antennas of diameter $D=12\,\rm{m}$. Therefore, we reduced the effective observing time ($t_{\rm{obs}}=12000\,\rm{s}$) by a factor of $f_{\rm{corr.}}=(50\cdot6^2)/8000^2=2.81\cdot10^{-5}$. We then fold the theoretical image with the final PSF of $15\,\rm{mas}$ to produce the images presented here.}
}
\end{itemize}

\section{Results}\label{sec:results}

\begin{figure*} 
	\begin{center}
		\includegraphics[scale=0.6]{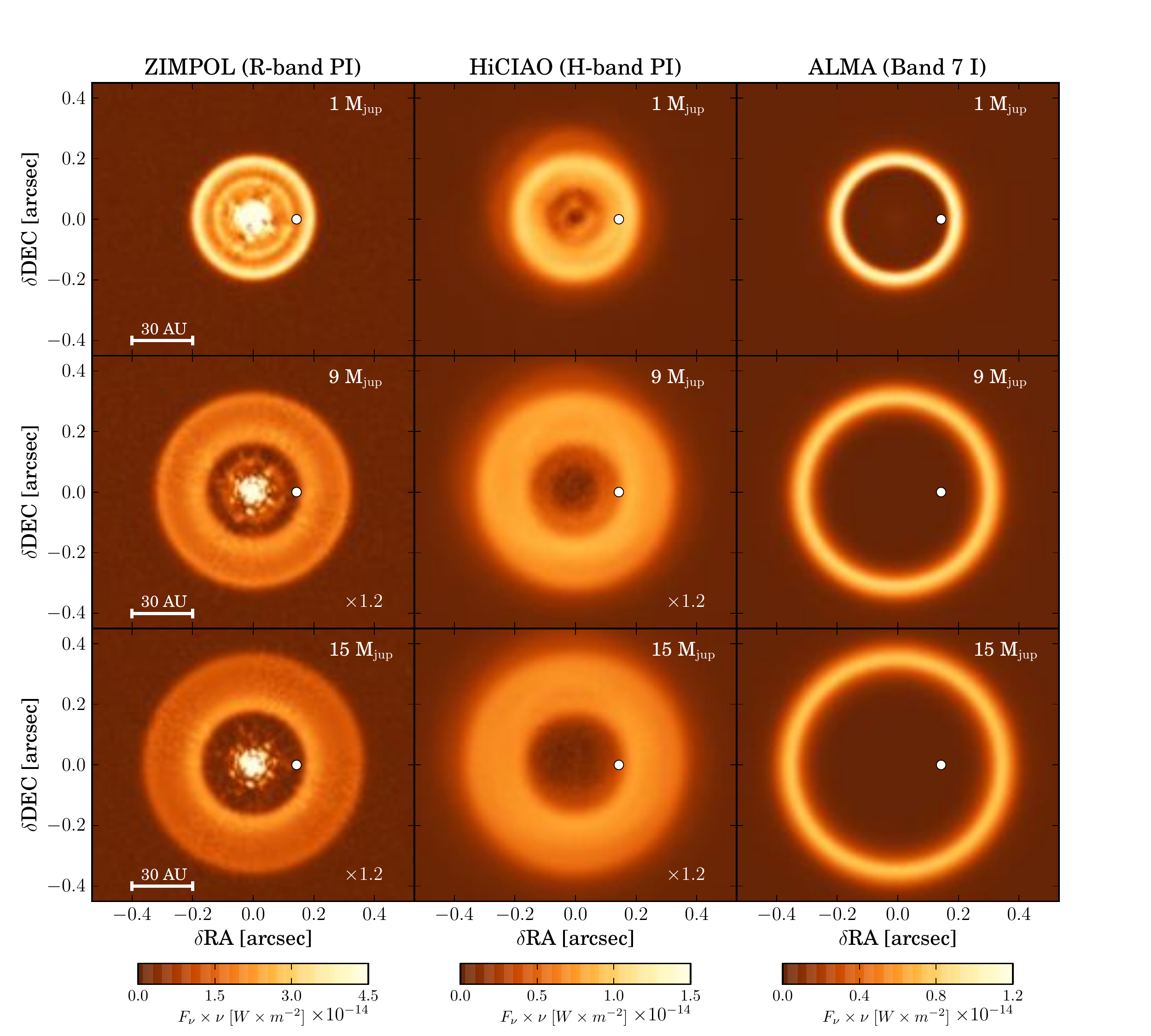}
	\end{center}
	 \caption{Simulated images of {the disc-planet models used in this study for the case of an embedded planet} orbiting at $20\,\rm{AU}$ (white dot). {Left to right columns correspond to polarised intensity (PI) ZIMPOL images in the {$R$}-band ($0.65\,\rm{\mu m}$), PI HiCIAO images in {$H$}-band ($1.6\,\rm{\mu m}$) and intensity (I) ALMA images at $850\,\rm{\mu m}$}. Top, middle and bottom rows show images obtained for planet masses of ${M_\mathrm{p}}=[1,9,15]\,{M_\mathrm{Jup}}$ respectively. All images in the same band (column) share the colour scale, although the cases of $9$ and $15\,{M_\mathrm{Jup}}$ in {$R$ and $H$} polarised intensity have been multiplied by a small factor to enhance the contrast. {See Section\,\ref{sec:method} for details on how the models/images were generated.}}
	 \label{fig:images}
\end{figure*}

\begin{figure*} 
	\begin{center}
		\includegraphics[scale=0.48]{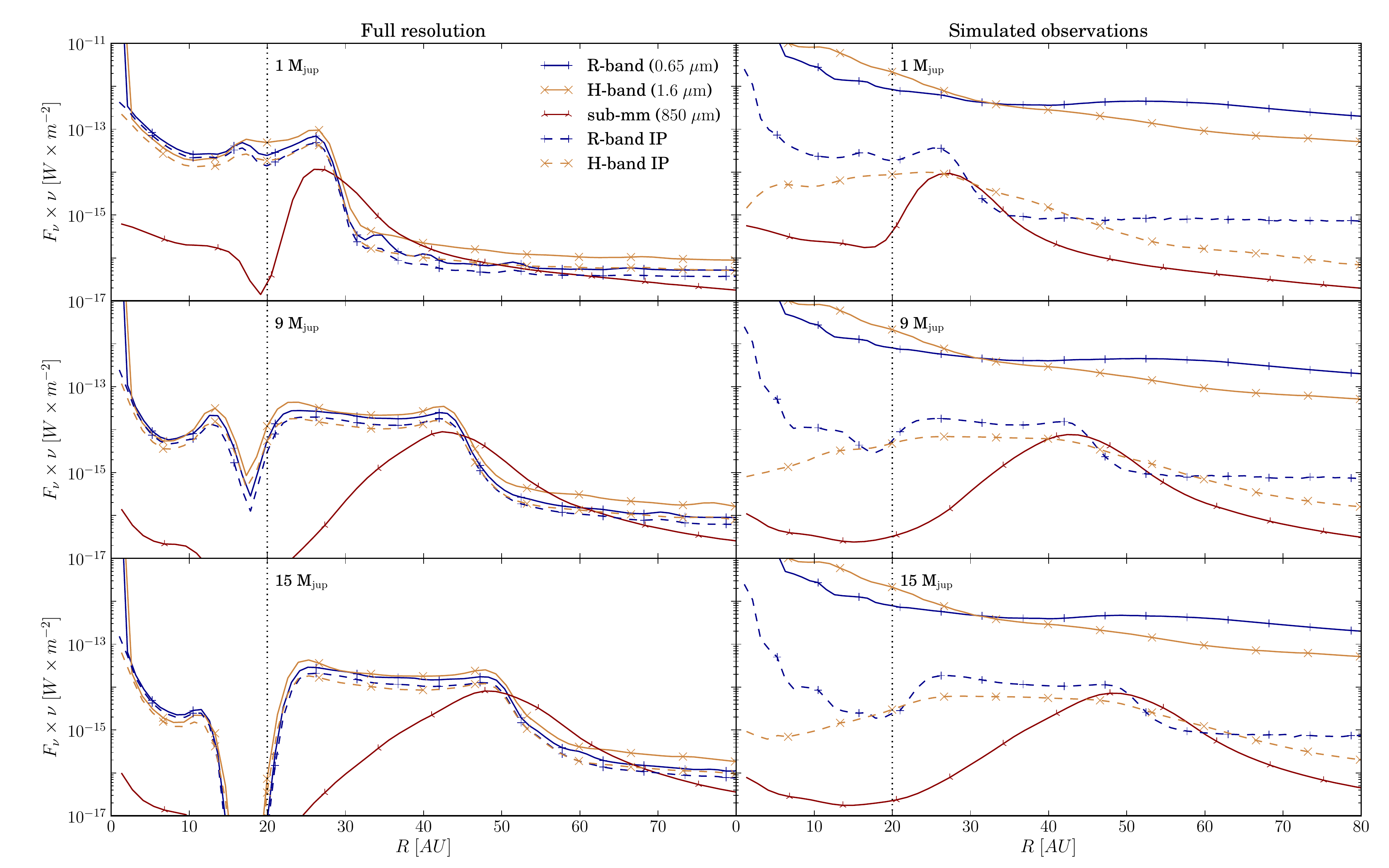}
	\end{center}
	 \caption{Radial profiles of the theoretical {images} (left panel) and simulated observations (right panel) of a $3\,\rm{Myr}$ disc with a planet of mass ${M_\mathrm{p}}=[1,9,15]\,{M_\mathrm{Jup}}$ from upper to lower panels, in the {$R$} ($0.65\,\rm{\mu m}$), {$H$} ($1.6\,\rm{\mu m}$) and $850\,\rm{\mu m}$ bands. Solid lines correspond to intensity ({I}) profiles while dashed lines correspond to polarised intensity ({PI}). The vertical dotted black line indicates the radial position of the planet at $20\,\rm{AU}$.}
	 \label{fig:radialprofiles}
\end{figure*} 

Figure\,\ref{fig:images} shows a selection of the obtained simulated observations for all planet mass cases at ${R_\mathrm{p}}=20\,\rm{AU}$ and after $3\,\rm{Myr}$ of evolution. Columns in the figure show, {from left to right, }polarised intensity ({PI}) for {$R$}-band with {ZIMPOL}, {$H$}-band {PI} with HiCIAO and intensity (I) with ALMA at $850\,\rm{\mu m}$, respectively. Rows show the different planet mass cases (i.e.\,${M_\mathrm{p}}=[1,9,15]\,{M_\mathrm{Jup}}$ from top to bottom, respectively), and the white dot represents the position of the planet at $20\,\rm{AU}$. Unless otherwise noted, radii lower than $R<{R_\mathrm{p}}$ or larger than $R>{R_\mathrm{p}}$ are referred to as \emph{inner} and \emph{outer} regions of the disc respectively.

{The polarised intensity images obtained with the polarimetric capabilities of ZIMPOL {(first column)} reveal} the structure of the disc {remarkably well}. In all planet mass cases, a gap at the position of the planet is detectable. Moreover, the three cases are different {because the inner regions are actually resolved}. In the case of polarimetric images obtained with HiCIAO and intensity images with ALMA, a gap of different sizes is also detected in all cases. However, these instruments do not detect any inner disc structure. {Note that we did not use a coronagraph in our simulations in order to get to the smallest possible working angles.  {Published observations executed with HiCIAO often do use a coronagraph, masking what amounts to the inner disk in our simulations {(see Sec.\,\ref{sec:discussion} below for details)}.}

Figure\,\ref{fig:radialprofiles} shows the computed intensity and polarised intensity radial profiles of the full resolution and (instrument) limited resolution images for all mass cases studied and a position of the planet of $R_{\rm{p}}=20\,\rm{AU}$. We show profiles for intensity in {$R$} ($0.65\,\rm{\mu m}$), {$H$} ($1.6\,\rm{\mu m}$) and $850\,\rm{\mu m}$ bands (solid lines), together with polarimetric intensity in {$R$} ($0.65\,\rm{\mu m}$) and {$H$} ($1.6\,\rm{\mu m}$) bands (dashed lines). Top to bottom panels show the cases of a planet mass of $M_{\rm{p}}=[1,9,15]\,{M_\mathrm{Jup}}$, {respectively}. Full resolution panels on the left, therefore show the theoretical emission and scattering pattern of the disc for each planet mass case, while the panels on the right (radial profiles of the simulated observations) show what would be imaged with the considered instruments (corresponding to the images in Figure\,\ref{fig:images}).

\subsection{Disc with a $1\,{M_\mathrm{Jup}}$ planet}\label{subsec:1Mjup}

The top row of Figure\,\ref{fig:images} shows the simulated observations with ZIMPOL, HiCIAO and ALMA for the $M_{\rm{p}}=1\,{M_\mathrm{Jup}}$ case at $R_{\rm{p}}=20\,\rm{AU}$. Both polarimetric images obtained with {ZIMPOL} in {$R$}-band and HiCIAO in {$H$}-band, detect scattered flux in {the $R<30\,\rm{AU}\,(\sim0.2\arcsec)$ region} that appears empty to ALMA at $850\,\rm{\mu m}$. However, the high spatial resolution delivered by {ZIMPOL}, allows to resolve a narrow and shallow gap at the position of the planet that is not detected by HiCIAO. 

The top left panel in Figure\,\ref{fig:radialprofiles} shows the radial profiles of the full resolution images for this planet mass case {(i.e.\,before applying the instrument simulators)}. In all these wavelengths, the morphology of the profiles can be divided into inner, {$r < 30\,\rm{AU}$}, and outer, $r>30\,\rm{AU}$, regions. Visible and {NIR} wavelengths show a sharp decrease in the scattered and emitted flux at this position ($r\sim30\,\rm{AU}$) {where} the emission at $850\,\rm{\mu m}$ presents a narrow ($\sim10\,\rm{AU}$) peak. In the inner $30\,\rm{AU}$ the scattered flux in both {$R$ and $H$} bands and the emission in {$R$} show the narrow depletion of about a factor of $\sim4$ at the position of the planet. The $850\,\rm{\mu m}$ emission profile is strongly depleted in this region.

{The top right panel in Figure\,\ref{fig:radialprofiles} shows the radial profiles of the simulated observations}. Emission profiles at visible and NIR wavelengths ({$R$ and $H$} bands) are completely dominated by the flux of the central star, thus the {structure of the disc seen in the full resolution {profiles is} lost in the convolution with the PSF of the instrument}. ALMA is immune to this effect {because the star is too faint at these wavelengths.} The high resolution of the instrument also allows for the detection of the ring at about $30\,\rm{AU}$. 

{Polarimetric observations in {$R$ and $H$} efficiently remove the stellar emission from the image and are able to show the inner edge of the outer disk.  However, because of the high resolution of the instrument, only ZIMPOL images in {$R$}-band show the local depletion at the position of the planet.}

These images and radial profiles are in agreement with what is expected from {the results of} \citet{pinilla12a}. The presence of the planet triggers the spatial separation of the different dust grain sizes (see bottom panels of Fig.\,\ref{2D_1D_Hydro_dust}). Big grains ($\sim 1\,\rm{mm}$) are trapped in the pressure maximum at about $\sim30\,\rm{AU}$, further out from the planet, and generate the emission detected by ALMA (i.e.\, ring at $\sim30\,\rm{AU}$). Small ($\sim 1\,\rm{\mu m}$) grains are allowed in the radii closer to the star ($R<30\,\rm{AU}$) and they are efficient scatterers at shorter wavelengths, which causes them to show up in the polarised intensity images.

\subsection{Variation with planet mass}

{For higher planet masses}, the spatial separation of dust grain sizes becomes stronger (see middle and right bottom panels of Fig.\,\ref{2D_1D_Hydro_dust}). The $9$ and $15\,{M_\mathrm{Jup}}$ planet mass cases are shown in the middle and lower panels of Figures\,\ref{fig:images} and \ref{fig:radialprofiles}. 

In both cases, the full resolution radial profiles show that both the outer sharp edge of the ring in visible and NIR wavelengths and the narrow emission ring at $850\,\rm{\mu m}$ are now located at {$\sim$$50\,\rm{AU}$}. The decrease in flux at the position of the planet becomes larger than one order of magnitude at all wavelengths. The gap becomes deeper and wider in all bands although for the sub-mm {wavelengths} this effect is stronger, indicating that indeed, big grains of dust are more subject to the depletion generated by the planet at these radii. 

The {loss of} structure in {$R$ and $H$} intensity images due to the {convolution with the instrumental PSF} is also present in these cases (blue and yellow solid lines in middle and lower right panels of Figure\,\ref{fig:radialprofiles}). Again, ALMA is able to detect the overall shape of the corresponding profile in the theoretical images, showing a ring at about $r\sim50\,\rm{AU}$ where the maximum of the theoretical profiles is placed.

The polarimetric images are also able to show structure in the $R<50\,\rm{AU}$ inner radii in these two planet mass cases. {$H$}-band {PI} images with HiCIAO, show an extended ring of scattered flux from $R\sim10\,\rm{AU}$ to $R\sim50\,\rm{AU}$ and from $R\sim20\,\rm{AU}$ to $R\sim50\,\rm{AU}$ for the intermediate and high planet mass cases respectively. 

The {ZIMPOL} images in {$R$}-band are particularly interesting in this cases. For the $9\,{M_\mathrm{Jup}}$ planet case, the simulated observation profile (dashed blue line in middle-right panel of Figure\,\ref{fig:radialprofiles}) shows an inner ring that extends from $R\sim5\,\rm{AU}$ to $R\sim15\,\rm{AU}$ that is not present in the more massive $15\,{M_\mathrm{Jup}}$ case. The presence of this inner ring is a very interesting feature that differentiates {between companion masses above or below the deuterium-burning limit of $\sim$13\,$M_\mathrm{Jup}$\footnote{{The deuterium-burning limit can range between {11--16}\,$M_\mathrm{Jup}$} depending on the metallicity \citep{spiegel11}}, which is often used as a dividing line between planets and brown dwarf companions.} {The full resolution radial profiles of the {$H$ and $R$} images in these two cases (blue and yellow lines in middle and bottom left panels of Fig.\,\ref{fig:radialprofiles}) show that the scattered flux from the inner $\leq17\,\rm{AU}$ radii in the $9\,{M_\mathrm{Jup}}$ case is about an order of magnitude higher than that of the $15\,{M_\mathrm{Jup}}$ planet} {in that same region}. Unfortunately, HiCIAO is not able to resolve it in the {$H$}-band{, and} in {ZIMPOL} images this radial region is dominated by {speckles} due to the proximity to the star. {Therefore, although the feature is resolved, its detection {is} {not reliable} enough}.

\subsection{Effect of the disc inclination}

\begin{figure*} 
	\begin{center}
		\includegraphics[scale=0.6]{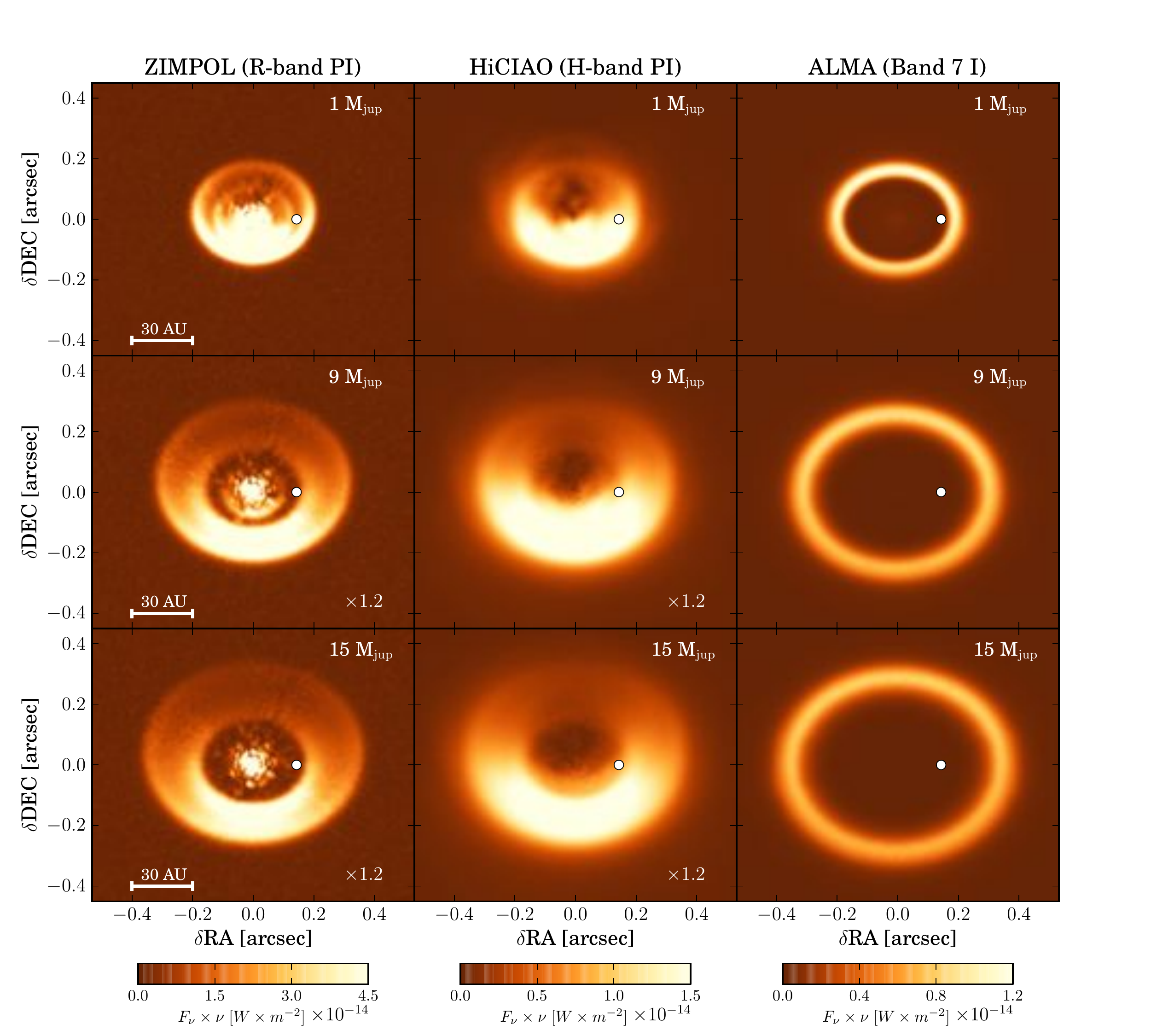}
	\end{center}
	 \caption{Simulated images of the same discs as the ones shown in Figure\,\ref{fig:images}, now with an inclination of $i=35^{\circ}$ \textbf(with respect to pole-on). Again columns, {from left to right, show {ZIMPOL} polarised intensity} in {$R$}-band ($0.65\,\rm{\mu m}$), HiCIAO {polarised} intensity in {$H$}-band ($1.6\,\rm{\mu m}$) and ALMA {Band\,7} ($850\,\rm{\mu m}$) intensity images, respectively. {Near and far sides of the disc correspond to lower and upper parts of the images, respectively.}} 
	 \label{fig:images35}
\end{figure*}

So far, we have considered the case of a disc with a $i=0^{\circ}$ inclination (i.e.\,pole-on) but the emission and scattering images of inclined discs can differ considerably from this case. Figure\,\ref{fig:images35} shows the same disc-planet cases presented in Figure\,\ref{fig:images} (i.e. $M_{\rm{p}}=[1,9,15]\,{M_\mathrm{Jup}}$ at $R_{\rm{p}}=20\,\rm{AU}$) with an inclination of $i=35^{\circ}$ {(angle measured from pole-on)}. In general, the structural features (i.e. gaps and rings) remain the same while the brightness patterns become asymmetrical. The polarised intensity images show clearly the effect of forward scattering, which makes the near side of the disc appear brighter than the far side. This effect is caused by the fact that dust particles, depending on their size with respect to the incoming wavelength, can scatter light differently in different directions. In particular, grains with sizes $2\pi a>\lambda$ are very strong forward scatterers which means that the population of $a>1\,\rm{\mu m}$ dust grains will show this effect in observations with {ZIMPOL} and HiCIAO {PI} at $\lambda\leq 1.6\,\rm{\mu m}$. The difference in brightness between both sides can be used as an estimator of the size of dust particles in the disc although one needs to be careful and appropriately account for the {effects of dust particle shape and structure} \citep{mulders13}.

In the transitional discs we model in this study, {the effect of the forward scattering} turns out to be even more useful since it makes possible to detect more clearly the dust particles in the inner regions ($R<R_{\rm{p}}$) of the disc. Indeed, the {ZIMPOL} images of the $35^{\circ}$ inclined discs for the $M_{\rm{p}}=9,15\,{M_\mathrm{Jup}}$ cases (middle and bottom images in the {first} column of Fig.\,\ref{fig:images35}) are now clearly distinguishable. The former shows scattering from the dust particles in the inner region of the disc while the latter does not, due to the fact that the depletion of particles is much higher in this case (see Section\,\ref{sec:discussion} below for details). 

The {ALMA simulations} show the opposite behaviour in the brightness pattern, where the far side appears brighter than the near side. This is simply due to the fact {that} we are observing the emission of warmer dust in the far side and colder dust in the near side. In other words, we are looking directly at the wall of {$1\,\rm{mm}$} particles illuminated by the star in the far side, while we are seeing the cold dust at the \emph{{``back''}} of the disc in the near side.

\subsection{Variation with planet position}

\begin{figure*} 
	\begin{center}
		\includegraphics[scale=0.6]{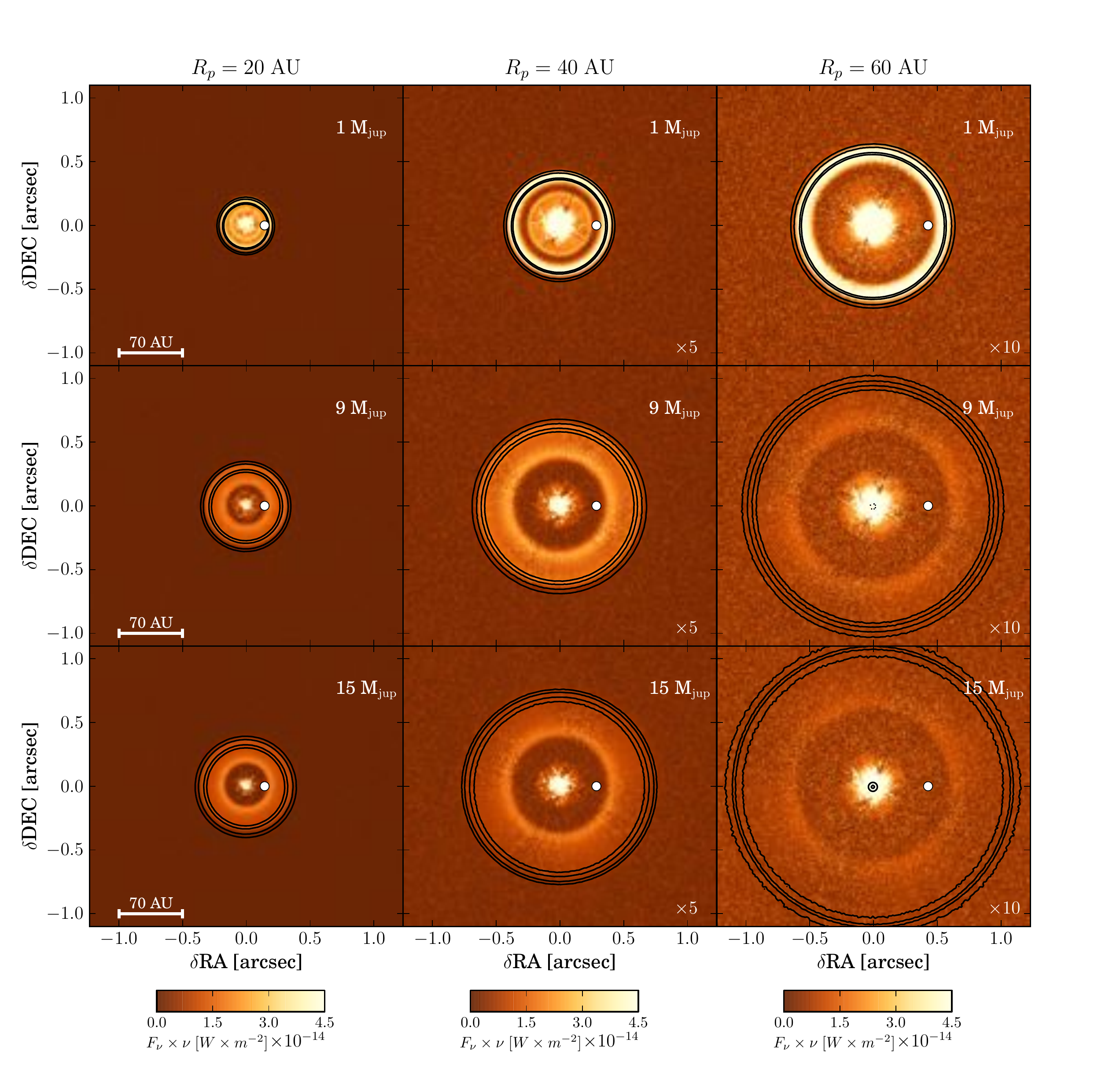}
	\end{center}
	 \caption{Simulated {ZIMPOL} images of the different mass and orbit radius cases studied. Upper to bottom rows correspond to a $3\,\rm{Myr}$ disc model with a planet of masses $M_{\rm{p}}=[1,9,15]\,{M_\mathrm{Jup}}$, respectively. Columns show, from left to right, the different planet orbit cases, $R_{\rm{p}}=[20,40,60]\,\rm{AU}$, respectively. The white dot indicates the position of the planet in each image and the black contour lines correspond to the intensity ring detected by ALMA at $850\,\rm{\mu m}$.}
	 \label{fig:plot4}
\end{figure*}

As the planet orbits further away from the star, the basic {inner disc--gap--outer disc} morphology of the disc remains the same for all planet mass cases, all re-scaled to the position of the planet. Figure\,\ref{fig:plot4} shows {ZIMPOL} images in {$R$-band} ($0.65\,\rm{\mu m}$) for all three cases of planet mass studied before, now at planet orbit radii of $R_{\rm{p}}=[20,40,60]\,\rm{AU}$. The white dot indicates again the position of the planet. {Overplotted}, the contour black lines show the emission {predicted for} ALMA at $850\,\rm{\mu m}$. 

The images show how the inner disc becomes more extended for larger orbit radii in all planet mass cases, although due to the larger distance between the dust and the star, the scattering weakens and the {ZIMPOL} images become fainter. This effect also increases with the mass of the planet in the outer part of the disc. 
{The radius of the ring detectable by ALMA (black contour lines), which traces the particle trap, increases in general with planet mass and separation.} 
It is important to note how for the case of $M_{\rm{p}}=1\,{M_\mathrm{Jup}}$, the gap in the {ZIMPOL} images remains quite narrow, tracing, almost exactly, the orbit of the planet {orbit radii of $20$ and $40\,\rm{AU}$. In the case of $R_{\rm{p}}=60\,\rm{AU}$ the inner disk becomes too faint to distinguish the outer edge}.

\subsection{Q-band $(20\,\rm{\mu m})$ measurements}

\begin{figure*} 
	\begin{center}
		\includegraphics[scale=0.6]{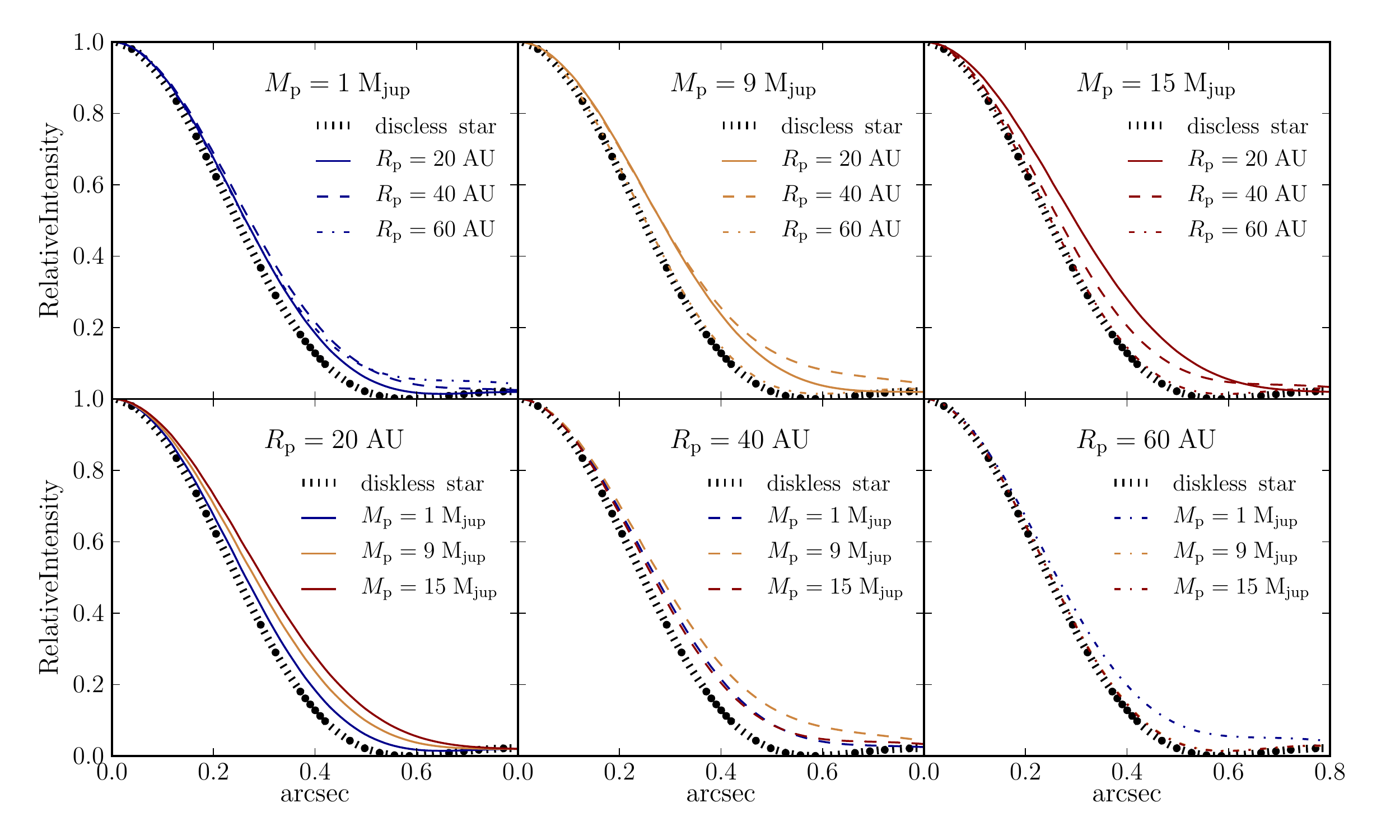}
	\end{center}
	 \caption{Synthetic $20\,\rm{\mu m}$ (Q-band) measurements for the different masses and planet positions considered. \emph{Upper row:} variation of the PSF profile with planet position for fixed planet mass. \emph{Lower row:} variation of the PSF profile with planet mass for a fixed planet position. {In all panels, the different colors indicate different planet masses (blue, yellow and red for $M_{\rm{p}}=[1,9,15]\,{M_\mathrm{Jup}}$, respectively) while different line styles correspond to different planet positions (solid, dashed and dash-dotted for $R_{\rm{p}}=[20,40,60]\,\rm{AU}$, respectively)}. The black {striped} line shows the PSF corresponding to a diskless (calibrator) star for reference.}
	 \label{fig:plot7}
\end{figure*}

{Mid-infrared emission (MIR) measurements generally trace warm $\sim10\,\rm{\mu m}$ dust emission. The presence of a circumstellar disc therefore broadens the footprint of the emission at {these} wavelengths and can cause the full width half maximum (FWHM) of the intensity profile to be larger than that of a point source. When a disc features an inner gap, the inner edge of the outer disc, i.e.\,the {``wall''}, is exposed to the radiation from the central star and its temperature increases, which increase the MIR emission from that radial position. The PSF of imaging instruments observing at these wavelengths is too large to resolve inner gaps (see Table\,\ref{tab:resolution}), but, if a wall is present, the FWHM of the intensity profile can be used, together with other diagnostics such as SED modeling, to estimate the position of the wall \citep{maaskant13}.}

{Figure\,\ref{fig:plot7} shows the variation of the intensity profile of our simulated observations at $20\,\rm{\mu m}$, for all masses and positions of the planets explored in this study. The {upper-row} panels show variation with position for a fixed mass while lower panels show variation with mass for a fixed position. In all panels, different colors correspond to different planet masses, different line styles to different planet positions and the profile of the calibrator star is shown as a {striped} black line for reference.}

The generally accepted idea is that the more exposed and the closer to the star the wall is, the more emission we expect. In this sense, for a fixed mass of the planet, the larger the semi-major axis the lower values we can expect for the FWHM. On the other hand for a fixed planet separation, a heavier planet depleting more mass inside the gap leaves the wall more exposed to the radiation and should therefore correspond to a larger values of the FWHM. However, one should be careful when making these correlations, because the warm dust in the inner disc (if present) also has a contribution that is, in fact, dominant in some cases, as our results show.

{Indeed, in Fig.\,\ref{fig:plot7} we see that when the position of the wall is relatively close to the star, {which holds for} all planet mass cases with semi-major axis at $20\,\rm{AU}$ (lower left panel), the \emph{``exposure''} effect dominates. More massive planets, which deplete more mass from the inner ($R\leq R_{\rm{p}}$) radii, yield broader intensity profiles. However, as the radial position of the planet increases the effect of the emission from the inner disc becomes increasingly important. At $R_{\rm{p}}=40\,\rm{AU}$ (lower middle panel), a $9\,{M_\mathrm{Jup}}$ planet will broaden the intensity profile more than a $15\,{M_\mathrm{Jup}}$. Since the wall is situated at roughly the same distance from the star (see images of Fig.\,\ref{fig:plot4}), this can only be attributed to the emission of a more massive (i.e.\,less depleted) inner disc. We can see the effect as well in the upper panels of Fig.\,\ref{fig:plot7}.}

{At $R_{\rm{p}}=60\,\rm{AU}$ (lower right panel) no {measurable} increase of the FWHM of the intensity profile is found for any planet mass. For all planet masses at this distance the inner disc is too depleted and the wall is too far away from the star (i.e.\,too cold) to contribute to the $20\,\rm{\mu m}$ emission.}

{Figure\,\ref{fig:SEDs} of Appendix\,\ref{ap:SEDs} shows, for completeness, {the SEDs} we obtained for all models presented in this study.}

\section{Discussion}\label{sec:discussion}

\begin{figure*} 
	\begin{center}
		\includegraphics[scale=0.5]{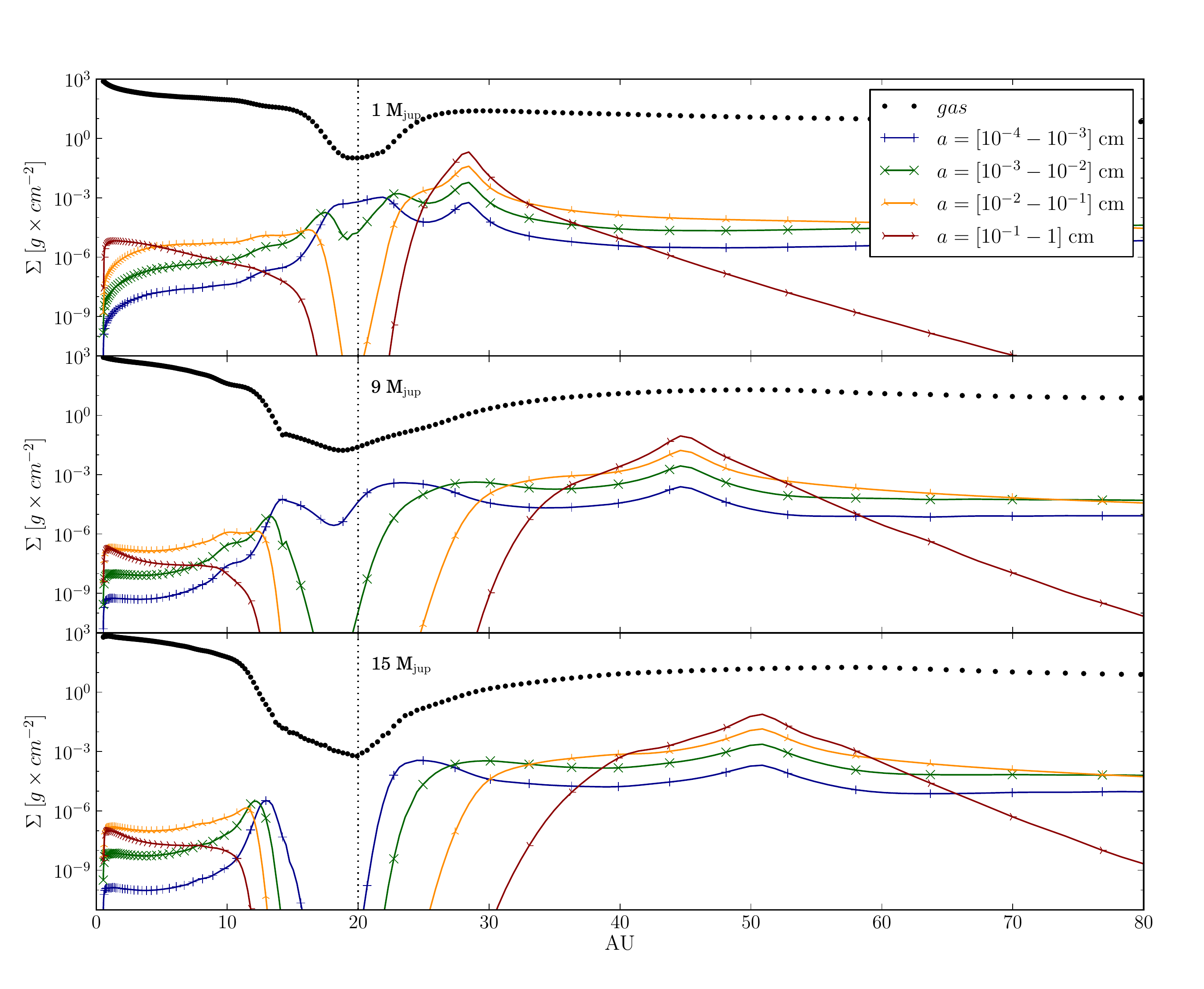}
	\end{center}
	 \caption{Surface density profiles for gas and different dust particle size ranges for the three planet mass cases, i.e.\,$M_{p}=[1,9,15]\,{M_\mathrm{Jup}}$. The vertical dotted line indicates the position of the planet at $20\,\rm{AU}$.}
	 \label{fig:plot3}
\end{figure*}

As explained in \citet{pinilla12a}, the presence of a massive planet affects the gas in the disc generating pressure gradients in {the pressure distribution, which would otherwise decrease monotonously with increasing separation}. If the disc is not perturbed by a planet, the negative gradient of the radial gas pressure makes the gas rotate with sub-Keplerian velocities while the dust moves with near-Keplerian velocities. This difference generates a drag force that causes the dust to lose momentum and spiral inwards \citep{weidenschilling77,brauer08a}. However, if a perturbation is present in the gas, whenever the pressure gradient is positive the gas velocity grows until it becomes Keplerian in a pressure maximum, which counteracts this inward radial drift causing the dust particles to accumulate in that position \citep{klahrhenning97,fromangnelson05,brauer08a,johansen09,pinilla12b}. What is interesting for our study is that the size of the particles that feel the trapping/filtering strongly depends on the strength of the pressure gradient \citep{pinilla12a} which depends mainly on the characteristics of the planet (i.e.\,its mass) and the turbulence parameter $\alpha$. In general, the bigger {a} particle, the {less susceptible it is to drag force, and the} more easily it {is trapped in the pressure maximum. Small particles, on the other hand, are able to filter through the pressure maximum and populate the inner regions} \citep{rice06}. 

As a consequence of these physical processes, the dust in the disc under the presence of a planet behaves, in general, as follows. Driven by the depletion of gas in radii close to the planet's orbit, the dust distribution undergoes a depletion that affects different dust grain sizes {in a different way}. Big {($\sim$$1\,\rm{mm}$)} dust grains are depleted more easily and ``pushed'' to radii where the pressure is maximum, further away form the planet. Small {($\sim$$1\,\rm{\mu m}$)} grains are able to remain at radii close to the planet{, ``escape'' the trap generated by the pressure bump, and flow} into inner regions due to the radial drift. As the planet becomes {more massive,} the depletion becomes stronger for all grain sizes, although the radial separation of different grain sizes remains. Figure\,\ref{fig:plot3} shows the surface density profiles of gas and different particle size bins in the studied planet mass cases {and planet orbit radius of $R_{\rm{p}}=20\,\rm{AU}$}. The solid blue line represents the surface density of very small particles {($a=[10^{-4}$--$10^{-3}]\,\rm{cm}$)}, while green, yellow and red represent particles of increasing size up to a maximum of $a=1\,\rm{cm}$. Although the simulations consider particles sizes up to {$2\,\rm{m}$,} the contribution of particles larger than $1\,\rm{cm}$ to the emitted and scattered light is {negligible,} so we did not include them in this plot for the sake of simplicity. According to the results presented in this study, the morphology of the dust distribution does not change with {the} planet's {orbital} radius but rather scales {with} it.

\subsection{Visible and near-infrared polarimetric images vs. sub-mm observations}
Since observations at different wavelengths trace (in general) different particle sizes, this spatial differentiation of the size distribution of dust in the disc opens the possibility to constrain planet parameters such as the mass and {separation} through multi-wavelength images of the emitted and/or scattered light from the disc. The images and radial profiles shown in Section\,\ref{sec:results} illustrate how the different observations trace different parts of this distribution. In general, {ZIMPOL} and HiCIAO polarised intensity {images} trace particles of about $1\to10\,\rm{\mu m}$ (blue solid line in Fig.\,{\ref{fig:plot3}),} because these particles are very efficient scatterers at these wavelengths while bigger grains are not. ALMA {Band\,7} ($\lambda\sim850\,\rm{\mu m}$) is very sensitive to emission from grains of about $1\,\rm{mm}$ and it therefore {highlights} the ring of big particles trapped in the pressure maximum, while the inner regions, heavily depleted of these grains, {appear} empty. 

ZIMPOL and HiCIAO images are {{different mainly} because of resolution}. Observing at shorter wavelengths, ZIMPOL is able to resolve the scattering due to the small grains in the inner region of the disc ({down} to the {separation} of the planet) while HiCIAO is only able to detect the outer part. It is important to note that, in our simulations, HiCIAO images do not include the remnants of the star in the center of the image, visible in ZIMPOL images. This is due to the fact that HiCIAO observations are simulated convolving polarimetric full resolution images that come out of MCMax with the measured PSF of the instrument at {$H$-band}. {Here we} assume that the star is totally unpolarised and that polarimetry is perfectly done which suppresses the starlight completely. ZIMPOL images are simulated more realistically using the ZIMPOL simulator which includes the speckle pattern that would remain in the observation due to imperfect polarimetry.



The presence of an inner disc in ZIMPOL images could directly differentiate {between a companion with mass below and above the deuterium-burning limit} since in the latter the depletion of small grains is so strong in these regions that the scattered flux is not detected. Moreover, a close look at the images of Figure\,\ref{fig:plot4} suggests a relation between the radial positions of the wall detected in ZIMPOL polarimetric images and the ALMA sub-mm emission peak at $850\,\rm{\mu m}$ (black contours) that varies with planet mass but {remains} approximately constant with planet {separation}. In order to investigate this further, we computed the ratio of these features for each case of planet mass and {separation} studied. 

Figure\,\ref{fig:plot6} shows the ratios versus planet mass for {planet orbit radii of $R_{\rm{p}}=[20, 40, 60]\,\rm{AU}$}. Using the radial profiles of the images, we define the position of the wall as the radial position of half the flux difference between the minimum flux at the bottom of the gap and the maximum flux at the wall (e.g.\,$23$ and $27\,\rm{AU}$ for minimum and maximum flux positions in the case of $1\,{M_\mathrm{Jup}}$ at $20\,\rm{AU}$, see the dashed blue line of upper right panel in Fig.\,\ref{fig:radialprofiles}). The upper and lower errors of the position are obtained {by} propagating the error on the determination of the mid-flux point and the peak of the emission in ZIMPOL and ALMA profiles respectively. Whenever those errors are lower than the corresponding resolution element ($4$ and $2\,\rm{AU}$, for ZIMPOL and ALMA respectively), the resolution element was used instead. The position of the points of same planet mass in the figure are slightly offset to facilitate their distinction.

The ratios for each planet's orbit radius are best fitted with a power law, 

\begin{equation}
f(M_{\rm{p}})=c\cdot \left(\frac{M_{\rm{p}}}{{M_\mathrm{Jup}}}\right)^{\gamma}\,, 
\end{equation}

with $c\sim0.85$ and $\gamma\sim[-0.22,-0.18,-0.16]$ for $R_{\rm{p}}=[20,40,60]\,\rm{AU}$ orbit radii, respectively. 

{Within the framework of our models, this figure serves} as a mass estimator for the planet.

\subsection{Mid-infrared vs. near-infrared polarimetric observations}

Interestingly, recent observational work presented {differences} also between mid-infrared and polarimetric observations. In a recent paper, \citet{maaskant13} presented the case of HD 97048 (a Herbig Be/Ae star with {a group} I flared disc) where $24.5\,\rm{\mu m}$ T-ReCS (Gemini South) measurements agreed with a model of the disc that features a gap of $\sim 30\,\rm{AU}$ (i.e.\,a wall positioned at $30\,\rm{AU}$). The reason why this is interesting is that this was the first study that suggested the presence of a gap in this disc, despite the fact that this target has been observed in multiple occasions at different wavelengths \citep{vanboekel04, lagage06,acke06a,doucet07,doering07,martinzaidi07,vanderplas09,carmona11,quanz12}. The closest-in observation, probing radial distances from {$\sim$$16$--$160\,\rm{AU}$}, was made by \citet{quanz12}{, who} took VLT/NACO {$H$- and $K_\mathrm{s}$-band} polarimetric images {in which} no gap was detected. 

\begin{figure} [htp]
	\begin{center}
		\includegraphics[scale=0.45]{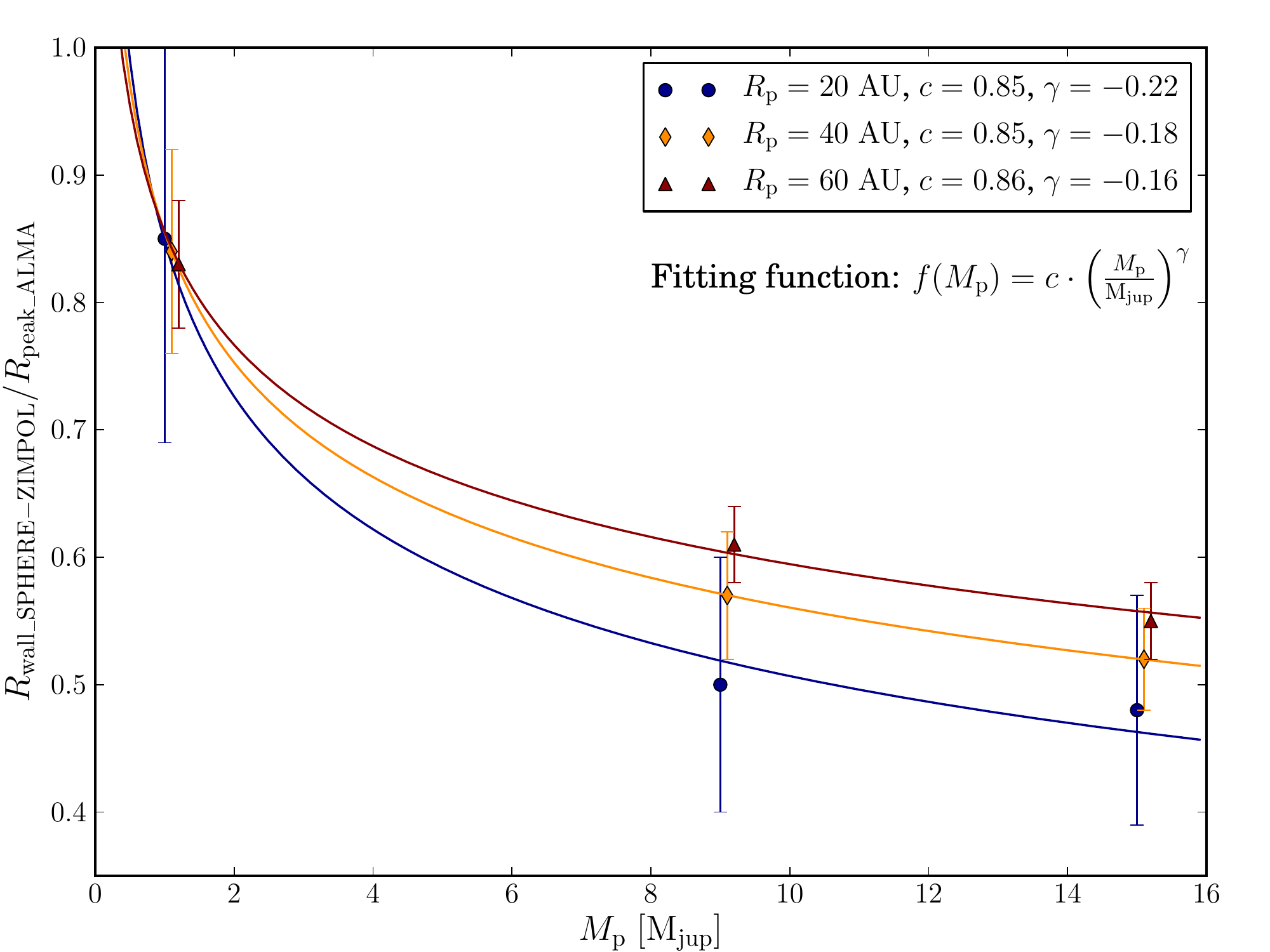}
	\end{center}
	 \caption{{SPHERE ZIMPOL} {$R$-band} outer gap edge to ALMA Band 7 peak ratio versus planet mass for the three planet orbit radii studied $R_{\rm{p}}=[20, 40, 60]\,\rm{AU}$.}
	 \label{fig:plot6}
\end{figure}

According to the results presented in this paper, these apparently discrepant observations would be consistent with the case of a disc hosting a planet of $1$ or $9\,{M_\mathrm{Jup}}$ at $20\,\rm{AU}$. Indeed the {$H$-band} Subaru/HiCIAO polarimetric images (similar to what would be observed with VLT/NACO) do not show a gap for these cases ({second} column upper panel in Fig.\,\ref{fig:images}), other than the inner $<10\,\rm{AU}$ {gap, }which can be attributed to the fact that we assume a perfectly suppressed central source (see Sec.\,\ref{subsec:instrcons} below). The $9\,{M_\mathrm{Jup}}$ {case, however,} shows a depletion in the inner $<20\,\rm{AU}$ that would have been most likely detected {by the \citet{quanz12} observations}. If, based on this, we discard the larger planet mass case, the upper left panel of Figure\,\ref{fig:plot7} shows how the PSF measured at {$20\,\rm{\mu m}$} would indeed appear broadened for the $1\,{M_\mathrm{Jup}}$ {case}. Here the cases of $1\,{M_\mathrm{Jup}}$ at $20\,\rm{AU}$, $1\,{M_\mathrm{Jup}}$ at $40\,\rm{AU}$ and $15\,{M_\mathrm{Jup}}$ at $40\,\rm{AU}$ (lower central panel) would all be consistent with the same broadening but, again, the fact that the polarimetric {$H$-band} images do not show a gap or a depletion of dust in these regions {discards} the two latter options. 

We note that the simulations presented here cannot be directly compared with either NACO or T-ReCS observations presented on those studies, {nor used as predictions for future observations of this target,} since the parameters of our star-disc system may differ considerably {from} those of HD 97048. Therefore we do not {claim} that a $1\,{M_\mathrm{Jup}}$ planet at $20\,\rm{AU}$ is indeed responsible for the observational diagnostics presented in those papers. We simply use {this observational} case as a proof of concept of our diagnostic method where polarimetric imaging at short wavelengths can be used to disentangle otherwise degenerate cases that would agree with measurements at larger wavelengths.
 

\subsection{Instrumental considerations}\label{subsec:instrcons}
Regarding instrumental performance, {and keeping in mind that the simulated images presented in this study for ZIMPOL and ALMA consider capabilities not available yet,} it is clear that, {whenever available, the upcoming ZIMPOL polarimeter} will {provide} unprecedented spatial resolution for polarimetric images. This is a big advantage since currently {available} polarimeters such as Subaru/HiCIAO or VLT/NACO do not have the power to probe the innermost regions of the disc as well as to accurately resolve the outer edge of the gap. According to the results of this study, these features are of extreme importance to properly constrain the mass and position of the planet. Although we are aware that our simulations have {limitations} (e.g. accretion of dust from the inner disc onto the star is not taken into account and we only consider the presence of one planet) and cannot represent all possible {disc-planet} systems, they serve as a proof of concept for the power of using combined interferometric and polarimetric images to characterise these objects. 

The results are in clear agreement with the {differences} observed between the SEEDS images obtained with HiCIAO and the interferometric results on the same targets. The gap shown by the ALMA images at $850\,\rm{\mu m}$ is larger than the gap in HiCIAO polarised intensity images at $1.6\,\rm{\mu m}$ by a factor {of $>$$2$} in radial extent, for all planet mass cases. If the disc is one of the $1\,{M_\mathrm{Jup}}$ type simulated here (or a less massive single planet), the gap in polarised intensity could be too small to even be detected. Moreover, if the polarimetric images are taken using a coronagraph, as it is the case for most of the SEEDS observations, this can hide the gap. From the disc-planet systems considered in our simulations, observations with HiCIAO in {$H$-band} using the coronagraph (${d_{\rm{eff.coro.}}\sim0.18\arcsec}$, \citealt{dong12}) would have missed all gaps generated by planets lighter than (and including) $9\,{M_\mathrm{Jup}}$. Although the basic effect has been shown in parametric models before, this is the first time that self-consistent physical models have been used to explain these observational discrepancies.
%

\section{Summary and conclusions}\label{sec:conclusions}

We present simulated imaging observations of a protoplanetary disc with a planet of masses $M_{\rm{p}}=[1,9,15]\,{M_\mathrm{Jup}}$ embedded in the disc and orbiting at $R_{\rm{p}}=[20, 40, 60]\,\rm{AU}$. We simulate intensity and polarised intensity images in the visible, near-infrared, mid-infrared and sub-mm wavelengths with current and near-future ground based imaging instruments with the aim of 1) test whether the proposed models explain observed {differences} between multi-wavelength observations recently presented, 2) study what different images can tell us about the dust distribution in the disc and 3) finding the best imaging strategy to infer planet mass and position. To simulate the disc-planet systems we use the models presented in \citet{pinilla12a} that combine 2-D hydrodynamical and state of the art dust evolution simulations to self-consistently compute the evolution of the dust and the gas in the system including radial drift, fragmentation and coagulation processes \citep{masset00,birnstiel10a}.

To simulate the observations, we first obtain the theoretical emission and scattered light images running the Monte-Carlo radiative transfer code MCMax on the gas and dust radial surface density distribution obtained from the \citet{pinilla12a} models after $3\,\rm{Myr}$ of evolution, for the six planet masses and positions studied. We then process the theoretical images obtained in different ways to simulate observations with {the upcoming} {SPHERE ZIMPOL} {polarimeter} in the {$R$-band} ($0.65\,\rm{\mu m}$ intensity and polarised intensity images), HiCIAO in the {$H$-band}  ($1.6\,\rm{\mu m}$ intensity and polarised intensity images), VLT-VISIR in the {$Q$}-band ($20\,\rm{\mu m}$) and ALMA ($850\,\rm{\mu m}$) with its {future} complete capabilities.

We find that:
\begin{enumerate}

\item{The trapping and filtering mechanisms triggered by the presence of a {$>$$1\,M_\mathrm{Jup}$ planet} in the disc {lead to different radial dust distributions for different grain sizes}. This causes observations at different wavelengths to show different structures. Particles with large sizes ($\sim1\,\rm{mm}$) are trapped in the pressure maximum outside the planet's orbit while smaller particles ($\sim1\,\rm{\mu m}$) are allowed to drift to the inner radii. The former show up in sub-mm emission measurements as a ring at the radial position of the pressure maximum, while the latter are detected in polarised scattering flux at inner radii.

The position of the pressure {``trap''}, the amount of particles trapped and filtered, and the particle size threshold of the filtering process depend strongly on the mass of the planet. This effect {allows} to constrain planet mass and {separation} by combining multi-wavelength observations.}\\
 
 \item{The \citet{pinilla12a} models are able to reproduce the {``missing cavities''} problem presented by the SEEDS survey ({\citealt{dong12}}; {$H$-band} images in this study), where {no gaps were found} in polarimetric {$H$-band} images for targets {known to exhibit gaps at $850\,\rm{mm}$}. Our simulated images with HiCIAO assume perfect polarimetry and a totally unpolarised star, which makes {it just} possible to detect a gap for all planet mass cases. In reality, speckle noise and, if used, the presence of a {coronagraph} would make the detection of the gap very unlikely.}\\
 
\item{Combination of sub-mm and polarimetric images in the visible wavelength {range} is the best imaging {strategy} to characterise the main features of the dust grain size distribution. An instrument like {SPHERE ZIMPOL} in the {$R$-band} ($\lambda=0.65\,\rm{\mu m}$) could differentiate between the three planet mass cases studied in polarised intensity, particularly if the system is inclined. The {high spatial resolution} provided by the instrument allows to resolve the different {structures in the inner region ({$[10$--$20]\,\rm{AU}$})}, or lack thereof, that each mass generates, which is currently not possible using any other instruments.}\\

\item{Combination of near-infrared polarimetric images and mid-infrared measurements can also yield constraints on mass and planet position, although less accurately than {the combination of diagnostics proposed as the best in this study} (visible polarimetric and sub-mm images). This is due to the fact that the {spatial} resolution of polarimetric images at these wavelengths ($1.6\,\rm{\mu m}$) is currently not enough to resolve the inner regions to the position of the planet. The warm dust in these regions also contributes to the FWHM measured at mid-infrared wavelengths. These causes some degeneracy in the models that would fit the measurements. Not having detailed spatial information in the polarimetric images for these regions (or even the gaps, which will be seen with {SPHERE ZIMPOL}) can make it difficult to disentangle the different planet mass and {separation} cases.}\\

 \item{{The results are also in agreement with the discrepancy between {$Q$}-band and near-infrared polarimetric observations of \citet{maaskant13} and \citet{quanz12} respectively, where a gap of {$\sim$$30\,\rm{AU}$} was found modeling the {$Q$}-band emission (at $24.5\,\rm{\mu m}$) while NACO polarimetric images in {$H$- and $K_\mathrm{s}$}-band did not show such {a} feature.}}

\end{enumerate}

\begin{acknowledgements}
The authors are grateful to Koen Maaskant, Gijs Mulders and the {``API cookie crew''} for helpful and insightful discussions, comments and suggestions during the course of the study. M.M. acknowledges funding from the EU FP7-2011 under Grant Agreement {No.\,284405.}{T.B. acknowledges support from NASA Origins of Solar Systems grant NNX12AJ04G.}
\end{acknowledgements}

\bibliographystyle{aa}

\bibliography{./mjsbib.bib}

\appendix
\section{Spectral energy distributions} \label{ap:SEDs}
\begin{figure*} [!htb]
	\begin{center}
		\includegraphics[scale=0.6]{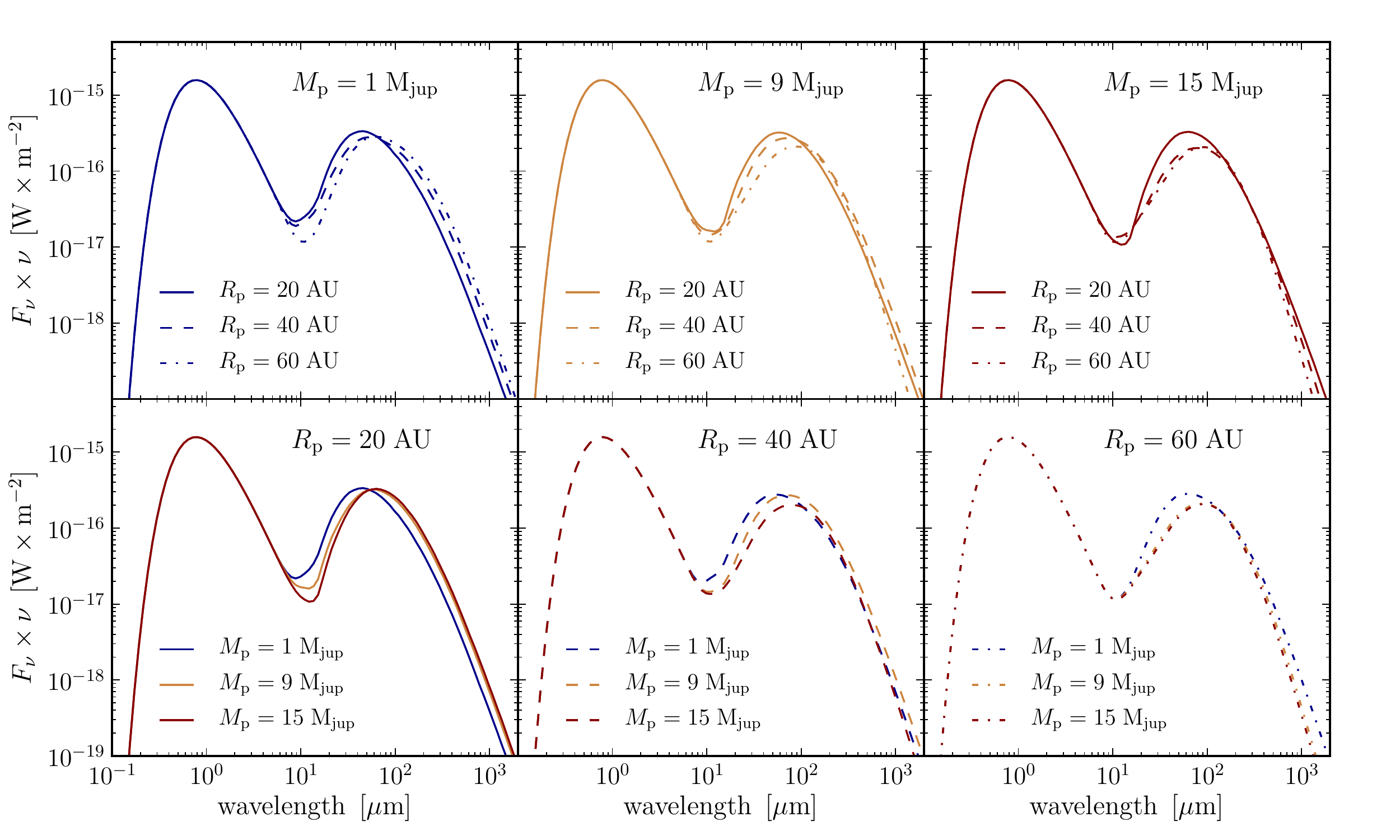}
	\end{center}
	 \caption{Synthetic Spectral Energy Distributions (SEDs) obtained for the different masses and planet positions studied. \emph{Upper row:} variation of the SED with planet position for fixed planet mass. \emph{Lower row:} variation of the SED with planet mass for a fixed planet position. In all panels, the different colors indicate different planet masses (blue, yellow and red for $M_{\rm{p}}=[1,9,15]\,{M_\mathrm{Jup}}$, respectively) while different line styles correspond to different planet positions (solid, dashed and dash-dotted for $R_{\rm{p}}=[20,40,60]\,\rm{AU}$, respectively)}
	 \label{fig:SEDs}
\end{figure*}

{Figure\,\ref{fig:SEDs} shows the Spectral Energy Distributions (SEDs) of the models used in this study. The format is the same as that of Figure\,\ref{fig:plot7} where upper panels show SED variation with planet separation for a fixed planet mass, and lower panels show variation with planet mass for a fixed planet separation.}

{The depletion of the SED in the range $\lambda\approx[20$ - $100]\,\rm{\mu m}$ increases systematically with both planet mass and separation. For longer wavelengths, in the far infrared (FIR) to millimetre regime (i.e. $\lambda\approx[100$ - $1000]\,\rm{\mu m}$), the flux increases with planet separation for the lowest planet mass case, whereas as the mass increases this relation seems to reverse. At the highest planet mass case, there is not significant variation with planet separation until $\lambda>1\,\rm{mm}$.}

{For fixed planet separations, the variation of the SED with mass in this region is only significant for the closest value, i.e\,$R_{\rm{p}}=20\,\rm{AU}$, although in the case of $R_{\rm{p}}=60\,\rm{AU}$ the lightest planet shows an excess in flux with respect to the $M_{\rm{p}}=9,15\,M_{\rm{Jup}}$ cases at long wavelengths.}

\end{document}